
%
%
%
%
%
%
\font\largebf =cmbx10 scaled\magstep1
\font\largeit =cmti10 scaled\magstep1
\newcount\Bcount \Bcount=0
\def\Beqn#1{\global\advance\Bcount by 1
  \global\xdef#1{\relax\the\Bcount} \eqno{\Bdis#1}}
\def\Bdis#1{\hbox{(B--#1)}}
\newcount\Acount \Acount=0
\def\Aeqn#1{\global\advance\Acount by 1
  \global\xdef#1{\relax\the\Acount} \eqno{\Adis#1}}
\def\Adis#1{\hbox{(A--#1)}}
\raggedbottom

\overfullrule=0pt

\def\frac#1/#2{\leavevmode\kern.1em
 \raise.5ex\hbox{\the\scriptfont0 #1}\kern-.1em
 /\kern-.15em\lower.25ex\hbox{\the\scriptfont0 #2}}
\def\fract#1/#2 {\leavevmode\kern.1em
 \raise.5ex\hbox{\the\scriptfont0 #1}\kern-.1em
 /\kern-.15em\lower.25ex\hbox{\the\scriptfont0 #2}}
 \baselineskip=16pt
\def\approx{\simeq}
%
%
\catcode`@=11
\newcount\chapternumber      \chapternumber=0
\newcount\sectionnumber      \sectionnumber=0
\newcount\equanumber         \equanumber=0
\let\chapterlabel=0
\newtoks\chapterstyle        \chapterstyle={\Number}
\newskip\chapterskip         \chapterskip=\bigskipamount
\newskip\sectionskip         \sectionskip=\medskipamount
\newskip\headskip            \headskip=8pt plus 3pt minus 3pt
\newdimen\chapterminspace    \chapterminspace=15pc
\newdimen\sectionminspace    \sectionminspace=10pc
\newdimen\referenceminspace  \referenceminspace=25pc
\def\chapterreset{\global\advance\chapternumber by 1
   \ifnum\the\equanumber<0 \else\global\equanumber=0\fi
   \sectionnumber=0 \makel@bel}
\def\makel@bel{\xdef\chapterlabel{%
\the\chapterstyle{\the\chapternumber}.}}
\def\sectionlabel{\number\sectionnumber \quad }
\def\unnumberedchapters{\let\makel@bel=\relax \let\chapterlabel=\relax
\let\sectionlabel=\relax \equanumber=-1 }
\def\eqname#1{\relax \ifnum\the\equanumber<0
     \xdef#1{{\rm(\number-\equanumber)}}\global\advance\equanumber by -1
    \else \global\advance\equanumber by 1
      \xdef#1{{\rm(\chapterlabel \number\equanumber)}} \fi}

\def\eqn#1{\eqno\eqname{#1}#1}

\def\eqinsert#1{\noalign{\dimen@=\prevdepth \nointerlineskip
   \setbox0=\hbox to\displaywidth{\hfil #1}
   \vbox to 0pt{\vss\hbox{$\!\box0\!$}\kern-0.5\baselineskip}
   \prevdepth=\dimen@}}
%

%

%

%
%
\newcount\fcount \fcount=0
\def\ref#1{\global\advance\fcount by 1
  \global\xdef#1{\relax\the\fcount}}
\def\refs{ \hangindent=10ex\hangafter=1}
\def\item{ \hangindent=4.4ex\hangafter=1}
%

%
%
\def\today{\number\year\space \ifcase\month\or 	January\or February\or
	March\or April\or May\or June\or July\or August\or September\or
	October\or November\or December\fi\space \number\day}
%
%
\def\spose#1{\hbox to 0pt{#1\hss}}
\def\simlt{\mathrel{\spose{\lower 3pt\hbox{$\mathchar"218$}}
     \raise 2.0pt\hbox{$\mathchar"13C$}}}
\def\simgt{\mathrel{\spose{\lower 3pt\hbox{$\mathchar"218$}}
     \raise 2.0pt\hbox{$\mathchar"13E$}}}
\def\simpropto{\mathrel{\spose{\lower 3pt\hbox{$\mathchar"218$}}
     \raise 2.0pt\hbox{$\propto$}}}
%
%
\newif\ifmathmode
\def\mathflag#1${\mathmodetrue#1\mathmodefalse$}
\everymath{\mathflag}
\def\displayflag#1$${\mathmodetrue#1\mathmodefalse$$}
\everydisplay{\displayflag}
\mathmodefalse

\font\tenmmib=cmmib10
\font\sevenmmib=cmmib10
\font\fivemmib=cmmib10
\newfam\bmitfam\textfont\bmitfam=\tenmmib
\scriptfont\bmitfam=\sevenmmib\scriptscriptfont\bmitfam=\fivemmib
\def\fixfam#1{\ifmathmode
                 {\ifnum\fam=\bffam
                    {\fam\bmitfam#1}\else
                    {\fam1#1}\fi}\else
                 {\ifnum\fam=\bffam
                    {$\fam\bmitfam#1$}\else
                    {$\fam1#1$}\fi}\fi}

\def\alpha{\fixfam{\mathchar"710B}}
\def\beta{\fixfam{\mathchar"710C}}
\def\gamma{\fixfam{\mathchar"710D}}
\def\delta{\fixfam{\mathchar"710E}}
\def\epsilon{\fixfam{\mathchar"710F}}
\def\zeta{\fixfam{\mathchar"7110}}
\def\eta{\fixfam{\mathchar"7111}}
\def\theta{\fixfam{\mathchar"7112}}
\def\iota{\fixfam{\mathchar"7113}}
\def\kappa{\fixfam{\mathchar"7114}}
\def\lambda{\fixfam{\mathchar"7115}}
\def\mu{\fixfam{\mathchar"7116}}
\def\nu{\fixfam{\mathchar"7117}}
\def\xi{\fixfam{\mathchar"7118}}
\def\pi{\fixfam{\mathchar"7119}}
\def\rho{\fixfam{\mathchar"711A}}
\def\sigma{\fixfam{\mathchar"711B}}
\def\tau{\fixfam{\mathchar"711C}}
\def\upsilon{\fixfam{\mathchar"711D}}
\def\phi{\fixfam{\mathchar"711E}}
\def\chi{\fixfam{\mathchar"711F}}
\def\psi{\fixfam{\mathchar"7120}}
\def\omega{\fixfam{\mathchar"7121}}
\def\varepsilon{\fixfam{\mathchar"7122}}
\def\vartheta{\fixfam{\mathchar"7123}}
\def\varpi{\fixfam{\mathchar"7124}}
\def\varrho{\fixfam{\mathchar"7125}}
\def\varsigma{\fixfam{\mathchar"7126}}
\def\varphi{\fixfam{\mathchar"7127}}

\unnumberedchapters


\font\fourteenbf=cmbx10 scaled\magstep2
\font\twelvebf=cmbx10 scaled\magstep1

 at 14.4truept
\font\namefont=cmr12
\font\addrfont=cmti12

\newbox\abstr
\def\abstract#1{\setbox\abstr=
    \vbox{\hsize 5.0truein{\par\noindent#1}}
    \centerline{ABSTRACT} \vskip12pt
    \hbox to \hsize{\hfill\box\abstr\hfill}}

\def\today{\ifcase\month\or
        January\or February\or March\or April\or May\or June\or
        July\or August\or September\or October\or November\or December\fi
        \space\number\day, \number\year}

\def\author#1{{\namefont\centerline{#1}}}
\def\addr#1{{\addrfont\centerline{#1}}}

\def\bx {{\bf x}}
\def\bk {{\bf k}}
\def\bg {{\bf \gamma}}
\def\bv {{\bf v}}
\def\E  {\eta}
\def\ie {{\it i.e.}}
\def\etal {{\it et al.}}
\def\eg {{\it e.g.}}
{ 
\nopagenumbers
{ 
{\hfill CfPA-94-Th-55; UTAP-193}
\vsize=9 truein
\hsize=6.5 truein
\raggedbottom
\baselineskip=18pt

\hfill
\hfill
\vskip3truecm
\centerline {
  \twelvebf {\fourteenbf T}OWARD
  {\fourteenbf U}NDERSTANDING {\fourteenbf CMB} {\fourteenbf A}NISOTROPIES}
	    }
\centerline {
  \twelvebf {\fourteenbf A}ND {\fourteenbf T}HEIR {\fourteenbf
I}MPLICATIONS\footnote{$^{\rm\dag}$}{\rm\negthinspace\negthinspace\negthinspace
 Submitted to PRD.  October 1994.  }
            }
\nobreak

\baselineskip=15pt
  \vskip 0.5truecm
  \author{Wayne Hu$^1$ and Naoshi Sugiyama$^{1,2}$}
	\smallskip
  \addr{$^1$Departments of Astronomy and Physics}
  \addr{and Center for Particle Astrophysics}
  \addr{University of California, Berkeley, California  94720}
  \vskip0.25truecm
  \addr{$^2$Department of Physics, Faculty of Science}
  \addr{The University of Tokyo, Tokyo, 113, Japan}
  \bigskip
  \bigskip
\noindent{\rm Working toward a model independent
understanding of cosmic microwave background (CMB) anisotropies
and their significance, we undertake a comprehensive and self-contained
study of scalar perturbation theory.
Initial conditions,
evolution, thermal history, matter content, background dynamics, and geometry
all play a role in determining the anisotropy.
By employing {\it analytic} techniques to illuminate the numerical results,
we  are able to separate and identify each contribution.
We thus bring out the nature
of the {\it total} Sachs-Wolfe effect,
acoustic oscillations, diffusion damping, Doppler shifts,
and reionization,
as well as their particular manifestation in
a critical, curvature, or cosmological constant dominated universe.
By studying the full angular {\it and} spatial
content of the resultant anisotropies, we isolate the signature
of these effects from the dependence on initial conditions.
Whereas structure in the Sachs-Wolfe anisotropy depends strongly
on the underlying power spectra,
the acoustic oscillations provide features which are nearly
model independent.  This may allow for future determination
of the matter content of the universe
as well as the adiabatic and/or isocurvature nature of the initial
fluctuations.

{\baselineskip=15pt \bigskip \smallskip
\vfill \noindent PACS: 98.70.Vc, 98.80.Es, 98.80.Hw. \hfil\break
hu@pac1.berkeley.edu, sugiyama@pac2.berkeley.edu}
} 
\eject
} 
\bigskip
{\bf \it  \baselineskip=12pt
\noindent It is only with people who know about the useless,

\noindent That there is any point in talking about uses.

\noindent In all the immensity of heaven and earth,

\noindent A man uses no more than is room for his feet.

\noindent Yet if in recognition we were to cut all else away,

\noindent Would it still be useful to man?

\smallskip
\qquad \qquad --Chuang-tzu}
\bigskip

\centerline{ \largebf I. Introduction}
\smallskip
With the steadily increasing number of cosmic microwave
background (CMB) anisotropy experiments on various angular scales
(see \eg\ [\ref\WSS\WSS]),
the empirical reconstruction of the process for structure
formation in the universe will soon enter a new phase.  For this task
to succeed, the groundwork for understanding anisotropy formation
must be firmly laid.
While numerical studies of
specific models abound, this {\it ab initio} black box
approach is not well suited to the reconstruction problem.
One must be able to distinguish between the effects of
initial conditions, evolution, thermal history, matter content,
background dynamics, and geometry.  With the goal of shedding light
on the model independent physical mechanisms
involved in anisotropy formation,
we have undertaken a comprehensive and self-contained study of the scalar
perturbations which give rise to large scale structure in
the universe.

Of the two general classes of scalar
 perturbations, the isocurvature
mode is by far the less well studied.  A rich structure of
anisotropies under the baryon isocurvature scenario is unveiled
by generalizing the original model proposed by Peebles [\ref\PeebPIB
\PeebPIB] to arbitrary thermal histories [\ref\EB\ref\Models\EB,\Models].
Yet even the familiar  adiabatic case
holds novel features if one steps beyond the standard $\Omega_0=1$
Harrison-Zel'dovich cold dark matter (CDM)
model [\ref\SugSilk \SugSilk].
In this paper, we extend the highly accurate analytic tools developed for
this standard CDM model [\ref\HSCDM \HSCDM]
to the general case of arbitrary initial conditions,
thermal history, and background dynamics.  By employing these methods
to illuminate the numerical results, we
examine the physical mechanism behind the
evolution of isocurvature and adiabatic fluctuations in
an $\Omega_0=1$, open, or cosmological constant dominated universe,
allowing for possible late or partial reionization.  Focusing on
the physical interpretation rather than specific model
dependent results, we explore the possibilities that these
as yet undetermined quantities leave open.

In \S II, we discuss the gauge invariant perturbation equations
and their general implications.
Unlike most previous analytic treatments, \eg\
[\ref\DSZ\DSZ, \ref\KS\KS], we take a {\it multifluid} approach
to realistically describe the evolution of each component.
Superhorizon evolution, analyzed
in \S III brings out
the differences between the isocurvature and
adiabatic modes, including the gravitational redshift [\ref\SW \SW]
and curvature effects [\ref\Wilson\Wilson].
Further discussion of open universe peculiarities may be found
in Appendix A, and commonly used relations in Appendix B.
As shown in \S IV,
intermediate scale perturbations in the photon-baryon fluid
evolve
as an oscillator in the potential well created
by the total density perturbations.  This leads to
the characteristic oscillatory ``Doppler'' peak structure in the CMB at
recombination for both modes [\ref\BEMNRAS\BEMNRAS].
Photon diffusion however erases these
acoustic oscillations at small scales [\ref\SilkDamp\SilkDamp].
This is especially
important in reionized scenarios, considered in \S V, where
last scattering is delayed and the diffusion length grows
to be nearly the horizon size at last scattering.  In this
case, degree scale anisotropies can be dominated by the Doppler
effect from scattering off electrons [\ref\SZDop\SZDop],
which at late times are released
from Compton drag.

Putting these results together in \S VI,
we examine their implications for the observable quantities
today.
By analyzing the full matter and temperature transfer functions,
we achieve separation of initial and evolutionary contributions.
Robust features in the anisotropy are singled out as potentially
useful for extracting information about the background cosmology.
We conclude in \S VII with some general comments on the present status
of models given their predictions for CMB anisotropies.
\bigskip
\goodbreak
{\centerline
{\largebf II. The Evolution Equations and Their Interpretation}}
\smallskip
In this treatment,
we assume scalar fluctuations about a background Friedmann-Robertson-Walker
metric,
$$
ds^2=\left[{a\over a_0}\right]^2 \left(-d\eta^2 + \gamma_{ij}dx^i dx^j \right),
\eqn\eqnFRW
$$
where $c=1$, $\gamma_{ij}$ is the 3-metric on a space of constant
negative curvature, and $\eta \equiv \int (a_0/a)dt$ is the conformal time.
Normalized to unity at matter-radiation equality,
the scale factor $a$ evolves as
$\dot a / a = Ha / a_0$ where the overdot denotes a conformal time derivative
and
$$
H^2 =
\left( {a_0 \over a }\right)^4 {1+a \over 1+a_0} \Omega_0 H_0^2
- \left( {a_0 \over a} \right)^2 K + {1 \over 3}\Lambda,
\eqn\eqnHubble
$$
is the Hubble parameter
with $H_0 = 100 h$ km s$^{-1}$ Mpc$^{-1}$ as its value today.
For spatially flat models, the
curvature
parameter $K = -H_0^2(1-\Omega_0-\Omega_\Lambda)$ goes zero, where
the vacuum density is related to the cosmological constant by
$\Omega_\Lambda = \Lambda/3H_0^2$.

Perturbations around these background quantities may be represented
in various ways under gauge invariant theory
\ref\Bardeen \ref\KSPert \ref\Mukhanov [\Bardeen,\KSPert,\Mukhanov].
Although all are gauge invariant,
they reduce to ordinary perturbation quantities
for different choices of hypersurface slicing [\ref\GSS\GSS].
This flexibility in
the gauge invariant scheme allows us to simplify
the physical interpretation.
For temperature perturbations, we choose shear free Newtonian slicing,
since on large scales, they are determined by
gravitational redshifts from the Newtonian potential.  Unfortunately,
this choice does not clearly bring out
the evolution of energy density perturbations, which is best studied
in the total matter rest frame representation.   To avoid
confusion, we will only employ photon and neutrino temperature perturbations,
$\Theta = \Delta T_\gamma/T_\gamma$ and $N = \Delta T_\nu/T_\nu$,
in the Newtonian representation, and energy density fluctuations, \eg\
$\Delta_\gamma =\delta \rho_\gamma/\rho_\gamma$ and
$\Delta_\nu =\delta \rho_\nu/\rho_\nu$, in the total matter
rest frame representation.

\bigskip
\goodbreak
\noindent {\largeit A. The Photon and Neutrino Boltzmann Equations }
\smallskip
The full linearized Boltzmann equation for
the evolution of the Newtonian photon
temperature perturbations $\Theta(\E,\bx,\bg)$,
is given by [\KS]
$$
\eqalign{
{d \over d\eta}(\Theta+\Psi)& \equiv
{\dot \Theta + \dot \Psi}  + \dot x^i {\partial \over {\partial x^i}}(\Theta +
\Psi)
+ {\dot \gamma^i} {\partial \over {\partial \gamma^i}}(\Theta + \Psi) \cr
& = \dot\Psi - \dot \Phi + {\dot \tau}(\Theta_0 - \Theta +
\gamma_i v^i_b + {1 \over 16}\gamma_i \gamma_j\Pi^{ij}_{\gamma}), \cr}
\eqn\eqnGen
$$
where ${\bf v}_b$ is the baryon velocity ($c=1$),
$\gamma_i(=\dot x_i)$ are
the direction cosines of the photon momentum, $\Theta_0$ is the
isotropic component of $\Theta$,
and the anisotropic stress perturbation for the photons
$\Pi^{ij}_{\gamma}$ is defined explicitly in Appendix A
from
the quadrupole moment of $\Theta$.
The last term in equation~\eqnGen\
accounts for Compton scattering, where $\dot \tau
= x_e n_e \sigma_T a/a_0$ is the differential optical depth,
with $x_e$ the ionization fraction, $n_e$ the electron number density,
and $\sigma_T$ the Thomson cross section.
The gauge invariant metric perturbations are $\Psi$, the Newtonian
potential and $\Phi$, the perturbation to the intrinsic spatial
curvature, which are related to the
density perturbation through a generalized Poisson equation in
\S IIC.
We will commonly refer to both $\Phi$ and $\Psi$ as gravitational
potentials.

If the potentials are static and Compton scattering is ineffective,
equation \eqnGen\ implies $\Theta+\Psi$
is a conserved quantity.  This merely represents what we call the
{\it ordinary} Sachs-Wolfe (SW) effect: a photon
experiences a fractional redshift of $\Psi$ climbing out of a $\Psi <0$
potential well.  The effective temperature perturbation accounting
for this shift is therefore $\Theta+\Psi$.
If $\Psi$ changes, the corresponding gravitational redshift of course
follows suit.  Changes in $\Phi$ also affect the photons through time dilation.
Since these effects accumulate along
the geodesics, we call the combination the {\it integrated} Sachs-Wolfe (ISW)
effect.  The {\it total} contribution, derived by
Sachs and Wolfe [\SW], is a combination of SW and ISW effects
and completely describes the effect of gravitational redshift on the photons.

In open universes, the $\dot \gamma_i$ term in equation~\eqnGen\
does not vanish due to the curving of geodesics.
Although this
would seem to complicate matters, its
effect on equation~\eqnGen\ is easy to interpret
and compute, once
we decompose the fluctuation into its normal modes.
Plane wave perturbations $Q =\exp(i\bk \cdot \bx)$, appropriate for a flat
geometry, must be replaced with the eigenfunctions of the Laplacian for an
open geometry \ref\Liftshitz \ref\Harrison \ref\AbSch [\Liftshitz, \Harrison,
\AbSch]:
$$
\nabla^2 Q \equiv \gamma^{ij}Q_{|ij} = -k^2 Q,
\eqn\eqnEigen
$$
where ``$|$" denotes a covariant derivative on the 3-space.
Since the eigenfunctions are complete
for $k \ge \sqrt{-K}$ one often introduces the
auxiliary variable $\tilde k^2 = k^2 + K$.  The subtle question of
whether $2\pi/k$ or
$2\pi/\tilde k$
should be considered as the ``physical'' wavelength of the mode
is examined
further in Appendix A.

Since each eigenmode evolves independently in linear theory, it is sufficient
to consider temperature
perturbations to exist in a single $k$-mode,\footnote*{As usual, the general
case can be recovered by summing a power spectrum of these
independent $k$-modes.
It should also be noted that all perturbation amplitudes such as
$\Theta_\ell$ have an implicit $k$-dependence.  However when discussing the
evolution of a single $k$-mode, we drop the index for brevity. After
this section,
no real space perturbation variables are employed.}
which can be decomposed into angular moments as
[\ref\Precursor \Wilson, \Precursor]
$$
\Theta(\eta,\bx,\bg) = \sum_{\ell=0}^{\infty} \Theta_\ell(\eta)G_\ell
(\bx,\bg).
\eqn\eqnLDecomposition
$$
Here the angular functions $G_\ell$
are defined in Appendix A such that they reduce to
$G_\ell = (-i)^{\ell} \exp(i \bk \cdot \bx) P_\ell (\bf k \cdot \bg)$
in the flat space limit, where $P_\ell$ is an ordinary Legendre polynomial.

We can now write equation~\eqnGen\ in the standard hierarchy of
coupled equations for the $\ell$-modes:
$$
\eqalign{
{\dot \Theta_0} &= -{k \over 3}\Theta_1 -{\dot \Phi},  \cr
{\dot \Theta_1} &= k\left[ \Theta_0 + \Psi -{2 \over 5}\left( 1 - {3K \over
k^2} \right)
\Theta_2 \right] - {\dot \tau}( {\Theta_1 - V_b}), \cr
{\dot \Theta_2} &= k \left[ {2 \over 3} \Theta_1
- {3 \over 7} \left( 1 - {8K \over k^2} \right)\Theta_3 \right]
- {9 \over 10} {\dot \tau \Theta_2}, \cr
{\dot \Theta_\ell} &= k \left[ {\ell \over 2\ell-1}\Theta_{\ell-1}
- {\ell+1 \over 2\ell+3}\left( 1 - \ell(\ell+2) {K \over k^2}\right)
\Theta_{\ell+1} \right] - {\dot \tau}\Theta_\ell, \quad (\ell > 2)
}
\eqn\eqnHierarchy
$$
where $\gamma_i v^i_b =  V_b G_1 = V_b (-k)^{-1} \gamma_i Q_{|i}$,
and we have made the replacements such as
$\Psi(\eta,\bx) = \Psi(\eta)G_0(\bx) = \Psi(\eta)Q(\bx)$ here and
below.
By analogy to equation \eqnHierarchy,
we can immediately write down the corresponding  Boltzmann equation
for (massless) neutrino temperature perturbations $N(\eta,\bx,\bg)$
by making the replacements
$
\Theta_\ell  \rightarrow N_\ell, \dot \tau \rightarrow 0,
$
in equation~\eqnHierarchy.
This is sufficient since neutrino decoupling occurs before any scale of
interest
enters the horizon.
\bigskip
\goodbreak
{\noindent \largeit B. CMB Anisotropies}
\smallskip
Although this Newtonian representation of the Boltzmann equation~\eqnHierarchy\
may cause stability problems for its numerical solution [\ref\NoteStable
\NoteStable],
it serves to bring out the physics of anisotropies quite well.
First, scattering tends to isotropize the photons in the electron
rest frame, leaving anisotropies only in the unscattered fraction:
for $\ell>2$,
$\Theta_\ell \propto
\exp(-\tau)$, whereas
$\Theta_2 \propto \exp(-9\tau/10)$
due to the angular dependence of
Compton scattering.  Isotropy also requires
$V_\gamma \equiv\Theta_1 = V_b$.
Even so, the dipole suffers from gravitational infall
due to $\Psi$, \ie\
the SW effect.
On the other hand, the ISW effect provides a source to the
monopole.

Since the density of free electrons decreases either due to
recombination, or if the universe is reionized, to the expansion,
the CMB eventually ceases to scatter
when the optical depth to the present from Compton scattering drops
to
$\int_{\eta_*}^{\eta_0}
\dot\tau d\eta =1$.
Under the standard recombination scenario,
this occurs at $z_* \approx 1000$, whereas for reionized
models it is delayed until
$$
z_* \approx 30 \left({\Omega_0h^2 \over 0.1}\right)^{1/3}
\left({0.05 \over x_e\Omega_b h^2} \right)^{2/3},
\eqn\eqnZLS
$$
if last scattering occurs before curvature or $\Lambda$ domination.

After $z_*$, the photons
effectively free stream to form anisotropies.
On the last scattering surface, the photon distribution may be
locally isotropic while still possessing inhomogeneities, \ie\
hot and cold spots, which will
be observed as anisotropies on the sky today.
Free streaming transfers fluctuations
to high multipoles, as the $\ell$-mode coupling of equation~\eqnHierarchy\
shows.  Consequently, in the absence of sources, the monopole
collisionlessly
damps.
For superhorizon  scales $k\eta \ll 1$, the photons
can only travel a small fraction of a wavelength, and thus the fluctuations
 remain in the monopole. This is reflected in the $k$-dependence of this
$\ell$-mode coupling.
If there is subsequent
reionization, superhorizon sized fluctuations will consequently
{\it not} damp by isotropization.

Due to the more rapid deviation of geodesics, a given length
scale will correspond to a smaller angle in an open universe than a flat
one.  Thus the {\it only} effect of negative
spatial curvature in equation~\eqnHierarchy\ is to speed the transfer of
power to higher multipoles.  Its effect is noticeable if the angular
scale $\theta \sim \ell^{-1}$ is less than the ratio of the physical
scale to the curvature radius $ \sqrt{-K}/k$.
One peculiarity arises however.  Even for the lowest eigenmode,
$k = \sqrt{-K}$ or $\tilde k = 0$, the $\ell$-mode coupling
in equation~\eqnHierarchy\ does not vanish.
Unlike the flat case, this ``infinite wavelength'' mode suffers
free streaming damping of low order multipoles,
once the horizon becomes larger than the curvature radius
$\eta\sqrt{-K} \simgt 1$.
The physical origin of this effect
is discussed further in \S VIB and Appendix A.

Finally, let us state some useful relations.
As discussed in Appendix A, the total anisotropy is
$$
\eqalign{
{2 \ell + 1 \over 4\pi} C_\ell &
= {V \over 2\pi^2} \int {d \tilde k \over \tilde k}
{M_\ell \over 2\ell + 1} \tilde k^3 |\Theta_\ell|^2 \cr
& = {V \over 2\pi^2} \int_{\sqrt{-K}}^{\infty}
{dk \over k}
{M_\ell \over 2\ell + 1} (1+K/k^2)^{1/2} k^3 |\Theta_\ell|^2,}
\eqn\eqnCl
$$
where the ensemble average anisotropy predicted for an experiment with
window function $W_\ell$ is $(\Delta T/T)^2 = \sum (2\ell+1)W_\ell
C_\ell/4\pi$ with $\Theta_\ell$ evaluated at present.
Here $M_\ell = (\tilde k^2  - K)...(\tilde k^2 - K\ell^2)
/(\tilde k^2 - K)^{\ell}$ and reduces
to unity in the flat space limit.
This implies that the contribution to the anisotropy
per logarithmic $k$ and $\ell$ interval is
$$
\left({\Delta T \over T}\right)_{\ell k}^2 \equiv
{\ell M_\ell \over 2\ell + 1}(1+K/k^2)^{1/2} k^3 V|\Theta_\ell|^2.
\eqn\eqnWeight
$$
We can also sum in $\ell$
to obtain
$$
|\Theta + \Psi|_{rms}^2 \equiv |\Theta_0 + \Psi|^2 +
\sum_{\ell=1}^{\infty} {M_\ell \over 2\ell+1}
|\Theta_\ell|^2,
\eqn\eqnRMSL
$$
which measures the total power in a single $k$-mode.
Since fluctuations are merely transferred to high multipoles by
free streaming, this quantity is conserved if $\dot \Phi = \dot \Psi =
\dot \tau = 0$,
as is evident from equation~\eqnGen.

\bigskip
\goodbreak
\noindent{\largeit C. Gravitational Potentials}
\smallskip
We of course have to define the gravitational potentials $\Phi$
and $\Psi$ in order to complete the Boltzmann equation~\eqnHierarchy.
It is useful to introduce the following gauge
invariant variables: the total density perturbation in the matter rest
frame
$\rho \Delta_T = \sum_i \rho_i \Delta_i$, where the sum
runs over all the species present and the density perturbations
are related to the temperature perturbations by
$$
\eqalign{
\Delta_\gamma &= 4\Theta_0 + 4{\dot a \over a}{V_T \over k}, \cr
\Delta_\nu &= 4N_0 + 4{\dot a \over a}{V_T \over k}; \cr}
\eqn\eqnDeltadef
$$
the total matter velocity $(\rho+p)V_T = \sum_i (\rho_i + p_i)V_i$,
where $p$ is the pressure;
and the
anisotropic stress $p\Pi = \sum_i p_i\Pi_i$, where contributions come
essentially from the radiation quadrupoles
$$
\Pi_\gamma = {12 \over 5} \Theta_2, \qquad
\Pi_\nu    = {12 \over 5} N_2.
\eqn\eqnAniso
$$
Employing the Einstein equations, we may now write the
generalized Poisson equation as
$$
\eqalign{
\Phi &= {4\pi G \over k^2 -3K}{\left( a \over a_0 \right)^2}
\rho\Delta_T, \cr
\Phi+\Psi &= -{8\pi G  \over k^2} {\left( a \over a_0 \right)^2}
p\Pi. \cr}
\eqn\eqnPotential
$$
As discussed above, scattering suppresses anisotropies
such that $\Theta_2 \approx 0$, and for perturbations larger
than the horizon scale, $N_2 \ll N_0$.  Moreover, in the matter
dominated regime, the pressure itself is negligible $p \ll \rho$.
The $p\Pi$ anisotropic stress term can thus  be
ignored as a first approximation,
implying $\Phi \approx -\Psi$.
\smallskip
\goodbreak
{\noindent \largeit D. Matter Components}
\smallskip
The baryons evolve under the generalized baryon continuity and Euler equations
$$
\eqalign{
\dot \Delta_b & = -k (V_b - V_\gamma) + {3 \over 4} \dot \Delta_\gamma, \cr
\dot V_b  &= -{\dot a \over a}V_b + k\Psi +
\dot \tau (V_\gamma - V_b)/R, \cr }
\eqn\eqnBaryon
$$
where $R \equiv 3\rho_b/4\rho_\gamma$ is the scale factor normalized
to ${\frac 3/4}$ at photon-baryon equality.
Again
if a collisionless non-relativistic particle were present, \eg\ CDM
or compact baryonic
objects [\ref\Gnedin\ref\COP \Gnedin,\COP], its evolution
would be obtained by setting $\dot \tau=0$.

Well inside the horizon,
equation~\eqnHierarchy\ and \eqnDeltadef\ imply that
the photons satisfy a separate continuity equation $\dot \Delta_\gamma
= -{\frac 4/3} k V_\gamma$, which reduces the first baryon equation
to the familiar form
$\dot \Delta_b = -kV_b$.
The baryon velocity decays due to the expansion and has a
source term from infall into gravitational wells.  Thus the only
effect of the decoupled components is through this potential term.

Early on scattering makes $V_b=V_\gamma$, which shows that the
photons and baryons evolve adiabatically $\dot \Delta_\gamma = {\frac 4/3}
\dot \Delta_b$, {\it regardless} of whether the initial conditions
are adiabatic or isocurvature.
Yet even if the universe remains fully ionized to the present, the
baryons will eventually decouple from the photons, because
$\dot \tau/R = {\frac 4/3} \dot \tau (\rho_\gamma / \rho_b)$
goes to zero in the matter dominated limit.
Since $\tau \propto \Omega_b$, this epoch is
independent of $\Omega_b$. The Compton
drag on an individual baryon does not depend on the total number
of baryons.  In fact,
equation~\eqnBaryon\ and the Poisson~\eqnPotential\ equations show that
the drag term $\propto V_b$ comes to dominate over
the gravitational infall term $\propto k\Psi$ at redshifts above $z \sim 200
(\Omega_0 h^2)_{\vphantom{e}}^{1/5} x_e^{-2/5}$.
Thus all modes are released from Compton drag at the same time,
which we take to be
$$
z_d = 160 (\Omega_0 h^2)_{\vphantom{e}}^{1/5} x_e^{-2/5},
\eqn\eqnZdrag
$$
defined as the epoch when fluctuations effectively join the
growing mode of pressureless linear theory (see \S VA).

It is important to realize that the drag and the last scattering redshift
are generally not equal.  The photons decouple from the
baryons {\it before} the baryons decouple from the photons
in the standard recombination scenario.
Typically the opposite occurs in reionized scenarios resulting in quite
different anisotropies for the two cases (see \S IV and V).

We now possess all the machinery necessary to describe the
evolution of perturbations.  Numerical solutions,
based on Sugiyama \& Gouda [\ref\SG \SG],
are presented in the following sections.  However, to shed light
on these solutions, we also apply analytic techniques
in the single fluid (\S III), photon-baryon fluid (\S IV), and
diffusive limits (\S V).

\smallskip
\goodbreak
\centerline {\largebf III. Large Scale Evolution: Single Fluid Approximation}
\smallskip
Since no causal process such as free streaming or diffusion
can separate the components, all fluid velocities are equal above
the horizon.
We can thus describe the coupled multi-component
system as a single fluid, defined by the total matter variables, whose
behavior does not even depend on the ionization history.
Its evolution is determined by combining the
equations for the various species, assumed to be either fully
relativistic or non-relativistic, \ie\ equations \eqnHierarchy\
and \eqnBaryon\ with their decoupled variants,
$$
\eqalign{
\dot \Delta_T - 3w {\dot a \over a} \Delta_T &=
-{\left( 1 - {3K \over k^2} \right)}(1+w)kV_T - 2 \left( 1 -{3K \over
k^2}\right)
{\dot a \over a} w \Pi, \cr
\dot V_T + {\dot a \over a}V_T &= {4 \over 3} {w \over (1+w)^2} k
[\Delta_T - (1-3w)S] +k\Psi - {2 \over 3}k
\left( 1 - {3K \over k^2}\right){w \over 1+w} \Pi,
 \cr}
\eqn\eqnTotal
$$
where $w = p/\rho$ and $S \equiv \Delta_m - {\frac 3/4}\Delta_r$,
with $\Delta_m$ and $\Delta_r$ being the perturbations in
the matter and radiation
energy densities
respectively.
As we shall see, $S$ can be interpreted as
an entropy fluctuation.
In the evolution equation for $V_T$,
infall due to the potential $\Psi$ is
countered by the pressure term $\Delta_T$ at small scales.
The  two are balanced at the Jeans scale.  In this section,
we solve the evolution equations neglecting pressure and
anisotropic stress
as is appropriate for large scales.

\smallskip
\goodbreak
\noindent{\largeit A. Initial Conditions and the General Solution}
\smallskip
The distinction between adiabatic and isocurvature scenarios
lies in the
entropy term $S$ of equation~\eqnTotal.  Its evolution is given by
the matter continuity equation \eqnBaryon,  \ie\
$\dot S = k(V_r - V_m)$, where the matter and radiation velocities
are defined in a manner analogous to $V_T$
[see equation~\eqnDeltadef].  Since all components have the same velocity,
$S$ is a constant before the mode enters the horizon and, if it is present,
must have been established at the initial conditions.

Under this $\dot S =0$ assumption,
we show in Appendix B that the
general growing solution of equation~\eqnTotal\ is
$$
\Delta_T = C_A U_A + C_I U_I,
\eqn\eqnGenSolution
$$
where the $C$'s are fixed by the initial conditions, and we have
neglected anisotropic stress.
Although $S$ is a constant above the horizon, we will define
$C_I \equiv S(0)$,
in anticipation of horizon crossing.
As we shall see in the next section, the evolutionary factors
$U_A$ and $U_I$ take simple asymptotic forms.  However to
preserve generality, we give the complete expressions here:
$$
\eqalign{
U_A & = \left[
      D^3
     + {2 \over 9} D^2
     - {8 \over 9}D
     -{16 \over 9}
     + {16 \over 9} \sqrt{D+1}   \right]
   {1 \over D(D+ 1)}\, , \cr
U_I &= {4 \over 15} \left( k \over k_{eq} \right)^2
	   \left( {1 - {3K \over k^2}} \right)
	   { 3D^2 + 22D + 24 + 4(4+3D)(1+D)^{1/2}
           \over (1+D)(4+ 3D)[1+(1+D)^{1/2}]^4 }D^3 }
\eqn\eqnDeltaLS
$$
respectively, where
$k_{eq} \equiv (a H)_{eq}/a_0
= \sqrt{2} (a_0 \Omega_0 H_0^2)^{1/2}$ corresponds to the scale
which passes the horizon at equality, and we have assumed $\Pi=0$.
The factor $D(a)$ accounts for pressureless growth
$$
D = {5 \over 4} k_{eq}^2 \tilde H \int {da/a_0 \over (\tilde Ha/a_0)^3},
\eqn\eqnDgrowth
$$
where $\tilde H$ is obtained from the Hubble parameter by ignoring the
radiation
$$
\tilde H^2 = { \left( a_0 \over a \right)^3 } \Omega_0 H_0^2 - \left(
{a_0 \over a} \right)^2 K + {1 \over 3}\Lambda.
\eqn\eqnReducedH
$$
We assume curvature and $\Lambda$ dynamical contributions are
only important well after equality $a\gg1$.
Curvature dominates over matter at $a/a_0 > \Omega_0/
(1-\Omega_0 -\Omega_\Lambda)$,
whereas $\Lambda$ dominates over matter at $a/a_0  >
(\Omega_0 / \Omega_\Lambda)^{1/3}$ and over curvature at
$a/a_0 > [(1-\Omega_0-\Omega_\Lambda)/\Omega_\Lambda]^{1/2}$.
Although we will usually only consider $\Lambda$ models which are flat,
these solutions are applicable to the general case.
Before either curvature or $\Lambda$ domination, $D = a$; afterwards,
it goes to a constant.
Moreover
if $a \gg 1$, $\Delta_T \propto D$
{\it regardless} of scale and initial
conditions as discussed below.

Whereas adiabatic scenarios begin with $C_I=0$ and finite $C_A$,
isocurvature universes have $C_I \ne 0$.
If only baryons, photons and neutrinos are present, as in the case of
the baryonic models, $S$ can be broken down into
$$
\eqalign{
S &= f_\nu \left(\Delta_b - {3 \over 4}\Delta_\nu \right)
	+ (1-f_\nu) \left(\Delta_b - {3 \over 4} \Delta_\gamma \right) \cr
       &= (1-f_\nu) S_{b\gamma} + f_\nu S_{b\nu}, }
\eqn\eqnEntropy
$$
with an obvious generalization to the case of an additional
decoupled non-relativistic component.  Here the neutrino fraction
$f_\nu = \rho_\nu/(\rho_\nu+\rho_\gamma)$ is time independent after
electron-positron annihilation, implying $f_\nu = 0.405$ for three
massless neutrinos and the standard
thermal history.  Since $S_{b\gamma} \equiv \Delta_b - {\frac 3/4}
\Delta_\gamma = \delta(n_b/n_\gamma)$,
and likewise for the neutrinos, $S_{b\gamma}$ and $S_{b\nu}$ represent
perturbations to the baryon to photon and baryon to neutrino number
densities respectively. This in fact justifies our use of the term ``entropy''
fluctuation to describe $S$.  Notice that $\dot S =0$ then has an obvious
interpretation: since the components cannot separate
above the horizon, the particle number ratios must remain constant.

Furthermore isocurvature conditions allow no initial
curvature perturbations as the name implies. Thus
the gravitational potentials $\Psi$ and $\Phi$ vanish at the initial
epoch, implying
$C_A = 0$.  If the perturbations were
formed in the radiation dominated epoch, they must be placed in the
baryons only. In this case, $S_{b\gamma}=S_{b\nu}=S$ initially,
 which we will hereafter adopt.
The generalization to the case where $S_{b\gamma} \ne S_{b\nu}$ is
straightforward.   Note that any arbitrary mixture of adiabatic and
isocurvature modes is also covered by equation~\eqnGenSolution.

\smallskip
\goodbreak
{\noindent \largeit B. The Evolution of the Fluid Components}
\smallskip
Now let us consider the implications and interpretation of
the general solution~\eqnGenSolution.  The results for the
adiabatic mode are extremely simple.
When the universe is dominated by radiation (RD), matter (MD), curvature
(CD) or the cosmological constant
 ($\Lambda$D), the total density fluctuation takes the form
$$
\Delta_T/C_A = {\cases { \vphantom{\big( }
			{10 \over 9} a^2 & RD \cr
		        \vphantom{\big( }
			a & MD \cr
		        \vphantom{\big( }
			D. & CD/$\Lambda$D \cr }}
\eqn\eqnDeltaA
$$
Moreover since $S=0$, the components evolve together
$\Delta_b = \Delta_c = {\frac 3/4} \Delta_\gamma = {\frac 3/4} \Delta_\nu$
where $\Delta_c$ is any decoupled non-relativistic component (\eg\ CDM).
The velocity
and potential are given by
$$
\eqalign{
V_T/C_A &= {\cases {\vphantom {\big( }
		 	-{5\sqrt{2} \over 12} (k_{eq}/k)(1 -3K/k^2)^{-1} a & RD \cr
		        \vphantom{\big( }
			-{\sqrt{2} \over 2}
			 (k_{eq}/k)(1-3K/k^2)^{-1} a^{1/2} & MD \cr
		        \vphantom{\big( }
			- (1-3K/k^2)^{-1} \dot D/k & CD/$\Lambda$D \cr }} \cr
\Psi/C_A &= {\cases {\vphantom {\big( }
		 	- {5 \over 6} (k_{eq}/k)^2 (1-3K/k^2)^{-1} & RD \cr
		        \vphantom{\big( }
			- {3 \over 4} (k_{eq}/k)^2 (1-3K/k^2)^{-1} & MD \cr
		        \vphantom{\big( }
			- {3 \over 4} (k_{eq}/k)^2 (1-3K/k^2)^{-1} D/a. &
			CD/$\Lambda$D  \cr }} \cr}
\eqn\eqnVPsiA
$$
Contrast this with the isocurvature evolution,
$$
\Delta_T/C_I = {\cases {
		   	\vphantom{\big( }
		\,{1 \over 6}\,( k /k_{eq} )^2 (1 - 3K/k^2)
			a^3
		   	\vphantom{\big( }
			& RD  \cr
		{4 \over 15} ( k / k_{eq} )^2 (1- 3K/k^2)
			a
		   	\vphantom{\big( }
			& MD  \cr
	 	{4 \over 15} ( k / k_{eq} )^2 (1- 3K/k^2) D.
		   	\vphantom{\big( }
			& CD/$\Lambda$D
		}}
\eqn\eqnDeltaLSCase
$$
{}From the definition of the entropy fluctuation $S$ (see also Appendix B),
equation~\eqnDeltaLS\
implies that
$$
\Delta_b/C_I = {\cases {
		   	\vphantom{\big( }
		1 - {3 \over 4}a & RD \cr
		   	\vphantom{\big( }
		{4 \over 3} \left[ a^{-1} + {1 \over 5}
		 ( k/k_{eq} )^2 (1 - 3K/k^2) a \right]
		   & MD \cr
		   	\vphantom{\big( }
		{4 \over 3} \left[ a^{-1} + {1 \over 5}
		 ( k/k_{eq} )^2 (1 - 3K/k^2) D \right],
		   & CD/$\Lambda$D \cr
		}}
\eqn\eqnDbLS
$$
and
$$
\Delta_\gamma/C_I = \Delta_\nu/C_I = {\cases {
		   	\vphantom{\big( }
		-a & RD \cr
		   	\vphantom{\big( }
		{4 \over 3} \left[ -1 + {4 \over 15}
		 ( k/k_{eq} )^2 (1- 3K/k^2) a \right]
		   & MD \cr
		   	\vphantom{\big( }
		{4 \over 3} \left[ -1 + {4 \over 15}
		 ( k/k_{eq} )^2 (1- 3K/k^2) D \right],
		   & CD/$\Lambda$D \cr }}
\eqn\eqnDrLS
$$
for the baryon and radiation components. Lastly, the velocity and the potential
also have simple asymptotic forms,
$$
\eqalign{
V_T/C_I &= {\cases {
		   	\vphantom{\big( }
		- {\sqrt 2 \over 8} (k/k_{eq}) a^2  & RD \cr
		   	\vphantom{\big( }
		- {2\sqrt 2 \over 15} (k/k_{eq}) a^{1/2}  & MD \cr
		   	\vphantom{\big( }
		- {4 \over 15} (k/k_{eq}) \dot D/k_{eq}  , & CD/$\Lambda$D \cr
}} \cr
\Psi/C_I &= {\cases {
		   	\vphantom{\big( }
		- {1 \over 8} a & RD \cr
		   	\vphantom{\big( }
		- {1 \over 5}  & MD \cr
		   	\vphantom{\big( }
		- {1 \over 5} D/a. & CD/$\Lambda$D \cr
}}
}
\eqn\eqnVPsiLS
$$
Notice that unlike the adiabatic case $V_T$ and $\Psi$ have no
explicit curvature
dependence in this representation.
Moreover, although these solutions omit radiation pressure and
streaming, they are actually valid for the {\it matter} all the way
to the present if horizon crossing occurs after the
drag epoch (see Fig.~1).

Let us try to interpret these results physically.
The isocurvature condition is satisfied by initially placing
the fluctuations in
the baryons $\Delta_b = C_I$ with
$\Delta_\gamma = 0$, so that $\Delta_T = 0$.
As the universe evolves however,
the relative significance of the baryon fluctuation
$\Delta_b \rho_b/\rho_T$ for the total density fluctuation
$\Delta_T$
 grows as $a$.  To compensate, the photon
and neutrino fluctuations grow to be equal and opposite
$\Delta_\gamma = \Delta_\nu = -a C_I$.
The tight coupling condition $\dot \Delta_b = {\frac 3/4}\dot
\Delta_\gamma$ implies then that
the baryon fluctuation must also decrease so that $\Delta_b
= (1-3a/4)C_I$.  The presence of $\Delta_\gamma$ means that there
is a gradient in the photon energy density.  This gradient gives rise
to a dipole $V_\gamma$ as the regions come into causal contact [see
equation~\eqnHierarchy],
\ie~$V_\gamma \propto k\eta \Delta_\gamma \propto -ka^2C_I$.
The same argument holds for the neutrinos.
Constant entropy requires that the total fluid move with the photons
and neutrinos $V_T = V_\gamma$, and thus infall,
produced by the gradient in the velocity, yields a
total density perturbation $\Delta_T \propto -k\eta(1-3K/k^2) V_T
\propto k^2(1-3K/k^2) a^3C_I$ [see equation~\eqnTotal].
This is one way of interpreting equation~\eqnDeltaLS\
and the fact that the entropy provides a source of total
density fluctuations in the radiation dominated epoch [\ref\Hu \Hu].

A similar analysis applies for adiabatic fluctuations, which begin instead
with finite potential $\Psi$.  Infall implies $V_T \propto
k\eta \Psi$, which then yields $\Delta_T \propto - k\eta V_T
\propto -k^2(1 - 3K/k^2) a^2 \Psi$,
thereby also keeping the potential constant.
Compared to the adiabatic case, the isocurvature scenario predicts
total density perturbations which are smaller by one factor of $a$
in the radiation dominated epoch
as might be expected from cancellation.

After radiation domination both modes
grow in pressureless linear theory $\Delta_T \propto D$
[{\it c.f.} equations~\eqnDeltaA\ and \eqnDeltaLSCase].
Whereas in the radiation dominated limit,
the entropy term $S$ and the gravitational infall term $\Psi$ are comparable
in equation~\eqnTotal, the entropy source is thereafter suppressed by
$w = p/\rho$, making the isocurvature and adiabatic evolutions identical.
Since growth is
suppressed in open and $\Lambda$ dominated universes, the potential $\Psi$
decays which has interesting consequences for anisotropies
as we shall now see.

\goodbreak
\smallskip
{\noindent \largeit C. The Total Sachs-Wolfe Effect: Basics}
\smallskip
As noted in \S IIB, the SW effect causes the effective
perturbation to be $\Theta+\Psi$ to account for the gravitational
redshift.  Subsequent changes in the potential of course
alter the shift, an effect which is approximately
doubled by the induced time dilation. This is the ISW contribution.
To determine the net effect however,
we must first derive the value of
the intrinsic photon fluctuations $\Theta$.  If $k\eta \ll 1$,
the Boltzmann equation~\eqnHierarchy\ reduces to the ISW effect
$$
\dot \Theta_0 = -\dot \Phi \approx \dot \Psi.
\eqn\eqnReduced
$$
Here we have again assumed $\Pi=0$, which causes a $\sim 10\%$
error.  For corrections due to $\Pi$ see [\HSCDM].

Since the isocurvature
initial conditions satisfy
$\Psi(0) = 0 = \Theta_0(0)$, this implies
$\Theta_0(\eta) = \Psi(\eta)$.
The effective
superhorizon scale temperature
perturbation for isocurvature fluctuations
is therefore
$$
\Theta_0 + \Psi = 2\Psi. \qquad ({\rm iso})
\eqn\eqnSWIso
$$
Inside potential wells,
the ISW effect makes
photons underdense
so that the gravitational redshift adds to the temperature perturbation.
This is a direct consequence of the feedback mechanism
which generates the potentials
(see \S IIIB).  Note however that in a low $\Omega_0h^2$ model
with standard
recombination, the potential may not reach its full
matter dominated value equation~\eqnVPsiLS\ by last scattering (see
Fig.~2).

For adiabatic perturbations, the nature of the growing mode $U_G$ [see
equation~\eqnDeltaLS] fixes
the initial perturbations to be
$
\Theta_0(0) = -{1 \over 2} \Psi(0),
$
reflecting the fact that the photons are overdense inside the
potential well.  Although $U_G$ implies the potential is constant in both
the matter and radiation dominated epoch, it changes to
$
\Psi(a) = {9 \over 10}\Psi(0)
$
through equality.
The ISW effect then brings the photon temperature perturbation in
the matter dominated epoch
to $\Theta(\eta) = - { 2 \over 3} \Psi(\eta)$ and the effective superhorizon
perturbation to (MD)
$$
[\Theta_0 + \Psi] = {1 \over 3} \Psi, \qquad ({\rm adi})
\eqn\eqnSWAdi
$$
which is the familiar Sachs-Wolfe result.

The above results are valid before last scattering at $a_*$.
Again since last scattering often occurs before full matter
domination, one should employ the full form
$[\Theta_0+\Psi](\eta_*) \approx \Theta_0(0)+2\Psi(\eta_*)-\Psi(0)$
instead of equations \eqnVPsiLS\ and \eqnSWAdi.
After $a_*$, the photons climb out of the potential wells,
leaving the quantity $[\Theta_0 + \Psi](\eta_*)$ to become anisotropies
today.  Additional contributions to the anisotropy may arise during
free streaming again through the ISW effect.  Since photon geodesics
are radial in the absence of scattering, we may use
the radial eigenfunctions of the Laplacian to solve for the free
streaming behavior.  These are denoted by
$Q(\bx)= X_\nu^\ell(\chi) Y_\ell^m(\theta,\phi)$, where $-\nu^2 = \tilde
k^2/K =k^2/K +1$, and the radial distance normalized to the
curvature radius is $\chi = \sqrt{-K}\eta$.
The radial function $X_\nu^\ell(\chi)$
reduces to $j_\ell(k\eta)$ in the flat limit.
For superhorizon scales at last scattering,
the resultant anisotropies are
(see Appendix A),
$$
{\Theta_\ell(\eta) \over 2\ell +1}
= [\Theta_0 + \Psi](\eta_*)
X_{\nu}^\ell(\chi-\chi_*)
+ \int_{\eta_*}^{\eta} (\dot \Psi - \dot \Phi)
X_{\nu}^\ell(\chi-\chi') d\eta'.
\eqn\eqnSachsWolfe
$$
The right hand side represents the SW and ISW effects respectively.
Since the potentials for both the adiabatic and isocurvature modes
are constant in the matter dominated epoch, the ISW contribution is
separated into two parts:
\smallskip
\noindent\item{(a)}
The early ISW effect due to pressure
growth suppression after
horizon crossing in the radiation dominated epoch.
\smallskip
\noindent\item{(b)}
The late ISW effect due to expansion growth
suppression
in the $\Lambda$ or curvature dominated epoch.

\noindent We shall now consider these effects in more detail.
\goodbreak
\smallskip
{\noindent \largeit D. The Total Sachs-Wolfe Effect: Detailed Structure}
\smallskip
Equation \eqnSachsWolfe\ for the total Sachs-Wolfe effect predicts a
rich structure of anisotropies for low $\Omega_0$ models [\SugSilk].
However, to build intuition for equation~\eqnSachsWolfe,
let us first
consider the familiar adiabatic $\Omega_0=1$ model in which the
ISW term represents only a small correction [\HSCDM].
A given $k$-mode
contributes maximum anisotropies to the angle that scale subtends
on the sky at last scattering.  In the $k-\ell$ plane, the anisotropy
will have a sharp ridge corresponding to this correlation (see Fig.~3).
Here we have plotted $(\Delta T/T)_{\ell k}^2$,
the logarithmic contribution to the anisotropy
in $k$ and $\ell$ defined by equation~\eqnWeight.\footnote*{For
 representation purposes, we chose the initial weights of
the $k$-modes to correspond to $|C_A|^2 \propto \tilde k$
and $|C_I|^2 \propto \tilde k^{-3}$ for the adiabatic and isocurvature
modes respectively.
This does not sacrifice generality
since one can easily scale the figure to an alternate
initial weighting.
Note all contour plots of the anisotropy represent the numerical results.}
The pure spherical
Bessel functions $j_\ell(x)$ show that
the series of ridges and filamentary structures
are due to the structure of the radial eigenfunctions
themselves.  Notice that the largest $k$-modes project onto the monopole
and do not contribute to anisotropies.

Now let us move onto the more complicated $\Lambda$ and open cases.
For $\Lambda$ models, the ISW term in equation \eqnSachsWolfe\
yields both early and late type contributions.
As we shall see (see \S IV), inside the horizon during the
radiation dominated era, the potential decays due to
pressure.
This leads to a significant early ISW term
which is projected onto a somewhat larger angle than the SW effect
itself since
it originates closer to the present. That this is present before
$\Lambda$ domination is clear from Fig.~4a.
Because the early ISW effect approaches its maximum of $[\Psi-\Phi](\eta_*)
\approx 2\Psi(\eta_*)$ and
the adiabatic SW effect is given by approximately ${\frac 1/3}\Psi(\eta_*)$,
the early ISW ridge is more
prominent than the SW ridge in adiabatic models.
However for scales that enter the horizon during matter domination,
the decay in the potential due to radiation is much less significant.
Thus the height of the second ridge drops significantly at larger
scales (see Fig.~4a).

After $\Lambda$ domination $a/a_0 \simgt
 (\Omega_0/\Omega_\Lambda)^{1/3}$,
the potential once again decays.  Note
however that for typical values of $\Omega_0 \simgt 0.1$,
this decay begins only
comparatively recently leading to late ISW contributions.
This has three significant consequences.

\noindent\item{(a)}
The largest $k$-modes contribute little to the
anisotropy due to the projection effect.  Notice that the late ISW
contribution intersects the $\ell =2$ edge of Fig.~4 at a smaller
scale than the SW effect.  In Fig.~5a, we plot the analytic
decomposition of contributions to a $k$-mode slice corresponding
to these large scales.  The smaller late ISW contribution in fact partially
cancels the SW effect.  Since the SW contribution has not undergone
free streaming oscillations at $\Lambda$ domination, the two effects
contribute coherently and cancel due to the decay of the potential.
\smallskip
\noindent\item{(b)}
Since the potential is still decaying at the
present, the late ISW effect can boost the low order multipoles
for all scales.  In Fig.~5b, we plot a smaller mode and show that the
late ISW effect is positive definite.  Recent contributions have
not free streamed to the oscillatory regime.  The ridge structure of
Fig.~4 is due to the late ISW effect adding with every other ridge
in the SW free streaming oscillation.
\smallskip
\noindent\item{(c)}
Contributions are
spread out over a time comparable to $\eta_0$.  As we shall see in
\S VB, this implies cancellation of the late ISW contribution
as the photon travels through many wavelengths of the perturbation
[\ref\HS\HS].
Thus late ISW contributions are rapidly
damped as the scale decreases leaving {\it only} those scales that project
onto large angles.
\smallskip
\noindent
Together these factors imply that if the $k$-modes are equally weighted
(scale invariant), the result will be
a rise toward low multipoles from the late ISW contributions [\ref\Kofman
\Kofman].  On the other hand if scales that are superhorizon sized at late
ISW generation are strongly weighted, there is a {\it relative suppression}
of low multipoles due to SW and ISW cancellation.

\ref\KamSper

Open adiabatic models follow similar physical principles yet still
yield significantly different anisotropies.
Both the late ISW contribution at large scales and
the
early ISW contribution at small scales contributes near the
maximum of
$2\Psi(\eta_*)$.  On most scales, the combined ISW
effect completely dominates over the SW contributions (see Fig.~6).
However, just as in the $\Lambda$ case, the late ISW contributions
boost the anisotropy in a larger angle than the SW effect for
a given $k$-mode.  Notice where the late ISW ridge intersects
$\ell=2$.  For the largest mode $k=\sqrt{-K}$, the SW effect
consequently can contribute mildly and cancel part of the late
ISW effect as in the $\Lambda$ case.
Moreover, this projection effect implies that at these
scales, the late ISW effect itself
is increasingly suppressed with $\ell$ (see Fig.~5c).
These curvature
scale contributions however
are suppressed by the cut off in the potential near
the curvature scale from the Poisson equation \eqnPotential.
This has the effect
of converting the SW ridge into a peaked structure and curves
the contours of the ISW ridge away from $k = \sqrt{-K}$. Of course
a change in the underlying power spectrum that weights the $k$-modes
can partially or completely remove this effect.

At smaller scales, the late ISW effect completely dominates the low order
multipoles (see Fig.~5d).
Finally notice the evolutionary effect of geodesic deviation.
Comparing Figs.~4 and 6, we see that
the fluctuations are more rapidly carried to high multipoles
than in the $\Lambda$ case.

Isocurvature models differ significantly
 in that the potentials {\it grow} until
full matter domination.  Strong early ISW contributions which
are qualitatively similar to the SW term will
occur {\it directly} after recombination and continue until full matter
domination (see Fig.~2).  Thus the projection
of scales onto angles will follow a continuous
sequence which merges the SW and
early ISW ridges (see Fig.~7).

For the $\Lambda$ case, the early ISW effect
completely dominates that of the late
ISW effect.  Thus the analytic separation shows that the total
ISW and SW effects
make morphologically similar contributions and the boost in low order
multipoles
is not manifest.  Moreover, the two add coherently creating a
greater total effect unlike the adiabatic case (see Fig.~5a,b).
 Open isocurvature models
behave similarly except that the late ISW contributions
near its maximum (late ISW ridge) is not negligible.
It is thus
similar to the adiabatic case ({\it c.f.} Fig.~5d and 8d) except that it
does not usually
dominate the {\it total} anisotropy.

Notice also that since
there is no curvature cutoff in the potentials,
the SW and early ISW ridge extends all the way to the largest mode
$k = \sqrt{-K}$.  The projection ridge intersects
$k=\sqrt{-K}$ at $\ell \approx 10$ which is the scale the ($\Omega_0=0.1$)
curvature
radius subtends at a distance $\eta_0$ (see Fig. 8a and Appendix A).
This is indicative of the fact that the lowest eigenmodes $k \rightarrow
\sqrt{-K}$, $\tilde k \rightarrow 0$ all contribute to
curvature scale fluctuations (see \S VIB and Appendix A).
\bigskip
\goodbreak
\centerline{ \largebf IV. Intermediate Scale Evolution:}
\centerline{ \largebf Acoustic Oscillations and Standard Recombination}
\smallskip
As the perturbation enters the horizon, we can no longer view the
system as a single fluid.  Decoupled components
such as the neutrinos free stream and change the entropy fluctuation.
However, above the photon diffusion scale, the photons and baryons are still
tightly coupled
until last scattering.  Since even at recombination, the diffusion length
is much
smaller than the horizon $\eta_*$, it is appropriate to combine
the photon and baryon fluids for study
[\ref\Jorgensen\Jorgensen,\ref\Fernando\Fernando,\ref\Uros \Uros].
In this section, we show that
photon pressure resists the gravitational
compression of the photon-baryon
fluid,
leading to {\it driven} acoustic oscillations
[\HSCDM]
\smallskip
\goodbreak
{\noindent \largeit A. The Acoustic or ``Doppler'' Peaks}
\smallskip
At intermediate scales, neither radiation pressure nor gravity
can be ignored. Fortunately, their effects can be analytically
separated and analyzed. Since photon-baryon tight coupling still
holds,
it is appropriate to expand the Boltzmann equation~\eqnHierarchy\
and the Euler equation~\eqnBaryon\ for the baryons
in the Compton scattering time $\dot \tau^{-1}$ \ref\PY [\PY].  To zeroth
order,
we obtain
$
\Theta_1 \equiv V_\gamma = V_b,
$
and $\Theta_\ell = 0$ for $\ell \ge 2$.
There is therefore no Doppler {\it shift} from
a scattering event, since the photons are already
isotropic in the baryon rest frame. Again this implies that the photon-baryon
evolution is adiabatic $\dot \Delta_b = {\frac 3/4}\dot \Delta_\gamma$
even when the general evolution is not.
The first order expansion yields
$$
\ddot \Theta_0 + {\dot R \over 1+R}\dot \Theta_0 +
k^2 c_s^2 \Theta_0 = F,
\eqn\eqnTight
$$
where the photon-baryon sound speed is
$$
c^2_s = {1 \over 3} {1 \over 1+R},
\eqn\eqnSoundSpeed
$$
and the forcing function is
$$
F = - \ddot \Phi - {\dot R \over 1+R}\dot
\Phi - {k^2 \over 3}\Psi.
\eqn\eqnForce
$$
The techniques established in our recent analysis of adiabatic
fluctuations in a flat universe [\HSCDM] work
equally well for the general case.  Under this formalism,
the gravitational driving forces are treated as known external potentials
in which the photon-baryon fluid oscillates.
The right hand side determines the effect of gravity through the ISW
effect and gravitational infall $k^2 \Psi$.
Notice that the SW
effect due to the photon's subsequent climb {\it out}
of the potential well partially counters infall [\HSCDM].
The ISW term $\ddot \Phi$ also drives the oscillation and is
important at horizon crossing
for modes that cross during radiation domination.  We shall see in \S IVC
that ISW contributions after last scattering can also counter or even
overwhelm the SW term.

The left hand side of equation~\eqnTight\ represents an oscillator
whose restoring force is due to the photon pressure.
This homogeneous
$F=0$ equation can be solved by
the WKB approximation,
$$
\eqalign{
\theta_a & = (1+R)^{-1/4} \cos kr_s, \cr
\theta_b & = (1+R)^{-1/4} \sin kr_s, \cr }
\eqn\eqnHomogeneous
$$
where the sound horizon is
$$
r_s = \int_0^\eta c_s  d\eta'
= {2 \over 3} {1 \over k_{eq}} \sqrt{6 \over R_{eq}}
  \ln { \sqrt{1+R} + \sqrt{ R + R_{eq} }
    \over
 	1 + \sqrt{ R_{eq}}},
\eqn\eqnSoundHorizon
$$
with $R_{eq} \equiv R(\eta_{eq})$.
Notice that if the sound speed is constant, the dispersion relation
becomes $\omega = kc_s$ as expected of acoustic oscillations.
The solution in the presence of the source $F$, constructed by
Green's method, is [\HSCDM]
$$
\eqalign{
[1+R(\eta)]^{1/4} \Theta_0(\eta) &= \Theta_0(0) \cos k r_s(\eta)
 + {\sqrt{3} \over k}
[\dot \Theta_0(0)+{1 \over 4}\dot R(0) \Theta_0(0)]\sin k r_s(\eta) \cr
&\qquad +
{\sqrt{3} \over k}
\int_0^\eta d\eta' [1+R(\eta')]^{3/4}
{\sin[k r_s(\eta)-k r_s(\eta')]} F(\eta')\, , \cr }
\eqn\eqnPartSoln
$$
and $
k\Theta_1 = -3 (\dot \Theta_0 + \dot \Phi).
$
Although the potentials in $F$ can be approximated from their
large (\S III) and small (\S IVB) scale solutions,
to show the true power of this
technique, we instead
employ their numerical values in Fig.~9.
The excellent agreement
with the full solution
indicates that our
technique is limited only by our knowledge of the potentials.
In almost all models, the potentials can at least be approximated
from the calculated matter power spectrum at the
present and the general principles of their evolution (see \eg\
[\HSCDM]).

Some common features of these acoustic oscillations, valid for
both isocurvature and adiabatic fluctuations are worthwhile to note.
At the start of the oscillation, the amplitude of the monopole increases
with $R$ (\ie\ $\Omega_b h^2$) due to a reduction in the pressure restoring
force.
Although both the monopole and the dipole subsequently decrease,
the dipole does so more rapidly due to an additional factor of
$\dot r_s = c_s \propto (1+R)^{-1/2}$.  Thus, when last scattering
freezes in the adiabatic oscillations, the temperature fluctuations
will be dominated by the monopole.  Furthermore, the amplitude of the monopole
is itself modulated since
inside a potential well,
the compressional phase of the oscillation is enhanced and the expansion
phase suppressed if $\Psi$ is comparable to $\Theta_0$ (see Fig.~9b).
This can also be viewed as a shift in the zero point of the oscillations
due to gravity. All even peaks for the adiabatic and odd peaks
for the isocurvature models suffer this suppression.
In models were the pressure is relatively low
(high $\Omega_b h^2$), the expansion
phase may be hidden entirely in the final anisotropy spectrum
[\HSCDM].

Those features which distinguish isocurvature from adiabatic
fluctuations are also apparent by inspection.
For adiabatic initial conditions, the driving potentials
are constant until Jeans crossing, at which point they decay (see Fig.~9).  On
the other hand, for the isocurvature scenario, they grow from zero
to a maximum at Jeans crossing.  Thus the forcing function imitates
$\cos kr_s$ and $\sin kr_s$
in the adiabatic and isocurvature scenario respectively
and stimulates the corresponding mode of temperature fluctuations.

\goodbreak
\smallskip
\noindent{\largeit B. From the Jeans to Diffusion Scale }
\smallskip
Well below the Jeans scale, the gravitational driving
force can be ignored and the photon-baryon fluctuations
behave as simple oscillatory functions,
until the breakdown of tight coupling at the photon diffusion
scale.  At this point, photon fluctuations
are exponentially damped due to diffusive mixing and rescattering.
We can account for this by expanding the Boltzmann and Euler equations
for the photons and baryons respectively to second order in $\dot \tau^{-1}$
\ref\PeeblesLSS
[\PeeblesLSS].
This gives the dispersion relation an imaginary part, making
the general solution
$$
\Theta_0 = A (1+R)^{-1/4}{\cal D}(\eta,k) \cos kr_s + B (1+R)^{-1/4}
{\cal D}(\eta, k) \sin kr_s ,
\eqn\eqnWKBSolution
$$
where the damping factor is
$$
{\cal D}(\eta,k) = e^{-(k/k_D)^2} ,
\eqn\eqnDamp
$$
with the damping scale
$$
k_D^{-2} = { 1 \over 6} \int d\eta {1 \over \dot \tau}
{R^2 + 4(1+R)/5 \over (1+R)^2} .
\eqn\eqnDampLength
$$
Diffusion thus dissipates these acoustic waves
leading to energy input and  spectral distortions in the CMB
 \ref\SZ \ref\HSSy [\SZ, \HSSy].

The amplitudes of these oscillations, \ie\
the constants $A$ and $B$, are determined by the total effect of
the gravitational driving force in equation~\eqnTight.
However, a simpler argument suffices for showing its general behavior.
As shown in \S IIIB, isocurvature fluctuations grow like
$\Delta_\gamma \approx -a C_I$ until Jeans crossing.
Since the Jeans crossing time is $a_J \sim k_{eq}/k$,
the isocurvature amplitude will be  suppressed by
$ k_{eq}/k $.
On the other hand, adiabatic fluctuations which grow as $a^2$
will have a $ (k_{eq}/k)^2$ suppression factor.
This simple argument fixes the amplitude up to a factor of order unity.

We obtain the specific amplitude by solving equation~\eqnTotal\
under
the constant entropy assumption $\dot S = 0$.
The latter approximation is not strictly
valid since free streaming of the neutrinos will change the entropy
fluctuation.  However, since the amplitude is fixed after
Jeans crossing,
which is only slightly after horizon
crossing, it suffices.  Under this assumption,
the equation can again be solved in the small scale limit.
Kodama \& Sasaki [\KS] find that for isocurvature fluctuations,
$$
A=0, \quad B= -{\sqrt{6} \over 4} {k_{eq} \over k} C_I, \qquad ({\rm iso})
\eqn\eqnAmpIso
$$
whereas for adiabatic perturbations,
$$
A={5 \over 4} \left( {k_{eq} \over k} \right)^2 C_A,
\quad B=0,  \qquad ({\rm adi})
\eqn\eqnAmpAdi
$$
if $k\gg k_{eq}$ and $k\eta \gg 1$.
As expected, the isocurvature mode stimulates the $\sin kr_s$ harmonic,
as opposed to $\cos kr_s$ for the adiabatic mode.

We can also construct the evolution of density perturbations
at small scales.  Well inside the horizon, $\Delta_\gamma = 4\Theta_0$
by equation~\eqnDeltadef.  The isocurvature mode solution therefore
satisfies (RD/MD)
$$
\Delta_\gamma/ C_I =  -\sqrt{6} \left( {k_{eq} \over k} \right) (1+R)^{-1/4}
{\cal D}(a,k) \sin kr_s.
\eqn\eqnDeltagSS
$$
The tight coupling limit implies $\dot \Delta_b
= {\frac 3/4} \dot \Delta_\gamma$ which requires (RD/MD),
$$
\Delta_b/ C_I = 1  -{3\sqrt{6} \over 4} \left( {k_{eq} \over k} \right)
(1+R)^{-1/4}
{\cal D}(a,k) \sin kr_s .
\eqn\eqnDeltabSS
$$
This diffusive suppression of the adiabatic component for the baryon
fluctuation
is known
as Silk damping [\SilkDamp].  After damping, the baryons are left with the
original entropy perturbation $C_I$.  Since they are surrounded by a
homogeneous and isotropic sea of photons, the baryons are unaffected by
further photon diffusion.
{}From the photon or baryon continuity equations at small scales,
 we obtain (RD/MD)
$$
V_b/C_I = V_\gamma/C_I \approx {3 \sqrt{2} \over 4} \left( {k_{eq} \over k}
\right)
(1+R)^{-3/4}
{\cal D}(a,k) \cos k r_s .
\eqn\eqnVSS
$$
As one would expect, the velocity oscillates $\pi /2 $
out of phase with, and increasingly suppressed compared to,
 the density perturbations.
Employing equations \eqnDeltagSS\ and \eqnDeltabSS, we construct
the total density perturbation by assuming that free streaming
has damped out the neutrino contribution
(RD/MD),
$$
\Delta_T/C_I = {a \over 1+a} \left[ 1 - {3\sqrt{6} \over 4}
	{k_{eq} \over k}
	 R^{-1}(1+R)^{3/4}
	{\cal D}(a,k) \sin kr_s
	\right] ,
\eqn\eqnDeltaSS
$$
{}From this equation, we may derive the potential
(RD/MD),
$$
\Psi/C_I = - {3 \over 4} \left( {k_{eq} \over k} \right)^2{1\over a}
\left[ 1 - {3\sqrt{6} \over 4}
{k_{eq} \over k}R^{-1}(1+R)^{3/4} {\cal D}(a,k) \sin kr_s
\right]  ,
\eqn\eqnVPsiSS
$$
which decays with the expansion since $\Delta_T$ goes to a
constant.
In Fig.~10, we compare these analytic approximations with the
numerical results. After damping eliminates the adiabatic
oscillations, the evolution of perturbations is governed by
diffusive processes.

A similar analysis for adiabatic perturbations shows that diffusion damping
completely eliminates small scale baryonic fluctuations.  Unlike
the isocurvature case, unless CDM wells are present
to reseed fluctuations (see \S V),
 adiabatic models consequently fail to form
galaxies. All adiabatic examples
shown here, including the open ones, are for CDM universes.

\goodbreak
\smallskip
\noindent{\largeit C. Recombination and Free Streaming }
\smallskip
If last scattering occurs before diffusion has damped the acoustic
oscillations in a given mode, \eg\ in the standard recombination
models, they will be frozen into the CMB.
A generalization of the free streaming equation of
\eqnSachsWolfe\ gives the resulting anisotropies,
$$
\eqalign{
{\Theta_\ell(\eta) \over 2\ell+1} &= [\Theta_0+\Psi](\eta_*)
	X_\nu^\ell(\chi-\chi_*) + \Theta_1(\eta_*)
	{1\over k}{d \over d \eta}X_\nu^\ell(\chi-\chi_*) \cr
	&\qquad + \int_{\eta_*}^{\eta} (\dot \Psi -\dot\Phi)
	X_\nu^\ell(\chi-\chi')
	d\eta' \, . }
\eqn\eqnFreeStream
$$
Here we obtain the diffusion damped fluctuation at last scattering
from equation~\eqnPartSoln\ by the replacements [\HSCDM]
$$
\eqalign{
[\Theta_0 + \Psi](\eta_*) &\rightarrow [\Theta_0+\Psi](\eta_*)
{\cal D}(\eta_*,k), \cr
\Theta_1(\eta_*) &\rightarrow \Theta_1(\eta_*)
{\cal D}(\eta_*,k), \cr }
\eqn\eqnDampCorr
$$
where
the damping factor is averaged over the finite duration of
last scattering
$$
{\cal D}(\eta_*,k) = \int d\eta \dot\tau e^{-\tau}
			e^{-(k/k_D)^2}.
\eqn\eqnDampLS
$$
Since the visibility function $\dot\tau e^{-\tau}$ goes to a delta function
for large $\tau$, this definition also coincides with
the tight coupling limit, equation~\eqnDamp.
For analytic approximations of the visibility function see
[\HSCDM,\ref\JW\JW].

As we have shown in [\HSCDM], equation \eqnFreeStream\ describes the
final anisotropies due to acoustic oscillations to high accuracy for
any given model.
However, for the task of reconstructing the model from observations,
it is useful to have a simple estimate of
equation~\eqnFreeStream.
As we have already seen with the Sachs-Wolfe effect, the presence of
the radial eigenfunctions $X_\nu^\ell$ in equation~\eqnFreeStream\
merely represents the projection of a spatial scale onto an angular
scale on the sky today.
The wavenumbers of the peaks in  the spectrum will correspond to
the modes in which
the monopole
reaches an extrema at last scattering: for adiabatic fluctuations
$k_m = m\pi/r_s(\eta_*)$, whereas for isocurvature fluctuations
$k_m = (m - {\frac 1/2})\pi/r_s(\eta_*)$, where $m$ is an integer $\ge 1$.
The first oscillation is thus at approximately the sound horizon at last
scattering $r_s(\eta_*)$.
This fluctuation is seen as an anisotropy in the multipole $\ell_m
\approx k_m r_\theta$, which corresponds to the angle subtended by the
scale $k_m$ at the distance of the last scattering surface.  In the
small angle approximation
$$
r_\theta = {1 \over \sqrt{-K}} \sinh[\chi_0-\chi_*]
\eqn\eqnRtheta
$$
which reduces to
$r_\theta = \eta_0 -\eta_*$ for a flat
universe.

The rapid deviation of geodesics in an open universe causes a
given scale to subtend a smaller angle on the sky.
In older universes (large present horizon
$\eta_0$ from $h$ or $\Lambda$), the distance to the last scattering surface
increases,
also reducing the angle a given scale subtends.  However,
this effect tends to cancel
with the corresponding increase in the sound horizon at last
scattering. The imperfect cancellation in the $\Lambda$ case
pushes the peak to somewhat smaller angles.

In Fig.~11, we plot the angular location $\ell_p = \pi r_\theta/r_s
(\eta_*)$ from which the series of peaks can be obtained as
described above.
This simple estimate does remarkably well
in tracing the higher peaks which are dominated by pure acoustic effects.
For the first peak, the potentials are still large enough at
last scattering to play a subsequent role. As mentioned in \S IVA,
the potential decays after horizon crossing due to radiation growth
suppression if the universe is not fully matter dominated.
For scales that cross the horizon before last scattering, this drives
the acoustic oscillations.  Afterwards it changes the gravitational
redshift that the photon would otherwise suffer from the SW effect.
We call the latter the {\it early}  ISW effect.
Since the sound horizon is always smaller than the particle horizon,
it contributes to scales between the first acoustic peak and
scales which cross the horizon during full matter domination.

For the
adiabatic mode, the
early ISW effect partially or fully
removes the SW redshift, uncovering the intrinsic fluctuation
$\Theta_0(\eta_*)$ from
the initial conditions and adiabatic growth,
as well as a contribution up to $\Phi(\eta_*)$
from the time dilation.  This can be seen in the additional contributions
to the rms temperature fluctuations after the fluctuation crosses the
horizon in Fig.~12a.
Since the potential has already
decayed by curvature or $\Lambda$ domination, further
decay does not significantly affect the radiation.
For scales
that cross well after radiation domination, the late
ISW effect is more significant and serves to
distinguish between $\Lambda$ and open models.  At intermediate scales, both
effects
are important.  However, since the late ISW contributions are
cancelled at small scales (see \S VB), the total anisotropy
is relatively smaller here.

For isocurvature models, the potential continues to grow outside
the horizon in the radiation dominated epoch (see Fig.~12b).
Therefore, evolution weights the large scales more heavily than
the small.  In fact, the first acoustic oscillation at last
scattering is not prominent due both to this enhancement of the
large scale and the continuing growth of the potential which
enhances the second oscillation.
This evolutionary tilt toward large scale anisotropies
can be countered
by changing the initial power spectrum.  However, the corresponding
enhancement of
small scale {\it matter} fluctuations makes reionization
likely in this model.
In this case, anisotropies are destroyed and then
regenerated by diffusive effects.
It is to this subject, we now turn.

\bigskip
\goodbreak
\centerline{\largebf V. Small Scale Evolution:}
\centerline{\largebf Diffusion Effects and Reionization}
\smallskip
Below the photon diffusion length, even photon-baryon tight coupling breaks
down.  Since the photons diffuse amongst the baryons, the two
fluids must be treated separately.
Moreover, if the universe is reionized, the diffusion length can grow
to nearly the horizon size at last scattering.
The window for adiabatic oscillations closes, and degree scale anisotropies
in the CMB will be dominated by diffusive effects.
Isocurvature baryon models also behave quite differently
from adiabatic
CDM models with respect to the matter.  For the decoupled CDM, density
perturbations grow regardless of ionization,
providing potential
wells into which the baryons may later fall.  Their absence
in the baryonic isocurvature case makes
the ionization history a crucial ingredient for structure formation
under this scenario.
Consequently although we retain generality for CMB anisotropies,
we concentrate on the isocurvature
model when discussing the effects of reionization on the matter.
\smallskip
\goodbreak
{\noindent \largeit A. Matter Evolution: Compton Drag}
\smallskip
{\noindent \largeit 1. Partial or Full Ionization}
\smallskip
Due to the lack of Silk damping,
baryon isocurvature models typically have high amplitude small scale
fluctuations which can collapse immediately after standard recombination
at $z \approx 1000$ [\PeebPIB].
It is possible that enough energy is released
to immediately reionize some fraction $x_e$ of the electrons.
This model will effectively behave as if recombination did not occur
at all.

Although the tight coupling approximation predicts $V_b$ and $\Delta_\gamma$
go to zero
inside the diffusion length, its breakdown keeps this from being
exactly satisfied ({\it c.f.} Fig.~10).
The single fluid Jeans argument of \S III become invalid. Since
in the diffusion limit where $S \approx \Delta_b \gg \Delta_\gamma$,
the effect of radiation pressure on $\Delta_T$
in equation~\eqnTotal\ is exactly canceled
by the entropy term.
After complete matter domination, the evolution equation and
its solution therefore becomes identical to the pressureless
case,
\ie\ all modes grow by the same factor $D(a)$ [see equation
\eqnDgrowth].

We can quantify this
with the Compton drag argument of \S IIC.  After $z_d  = 160
(\Omega_0 h^2)_{\vphantom{e}}^{-1/5} x_e^{-2/5}$, the baryons
are effectively released from photon pressure.
Thus, perturbations will grow such that
$\Delta_T(a) = [D(a)/D(a_d)] C_I$ for $a \gg a_d $.
An excellent empirical approximation (see Fig.~13)
to the behavior at intermediate times is given by
$$
\Delta_b/C_I = {\cal G}(a,a_d),
$$
with the interpolation function
$$
{\cal G}(a_1,a_2) = 1 + {D(a_1)\over D(a_2)}\exp(-a_2/a_1),
\eqn\eqnDragGrow
$$
where if $a_1 \gg a_2$, ${\cal G}(a_1,a_2) \rightarrow D(a_1)/D(a_2)$.
The velocity $V_T$ is given by the continuity equation~\eqnTotal.
Notice that this properly accounts for growth in
an open and/or $\Lambda$ universe.
\smallskip
\goodbreak
\noindent{\largeit 2. Late Ionization}
\smallskip
Now let us consider more complicated thermal histories.
Standard recombination may be followed by a significant
transparent period before reionization at $z_i$, due to some later
round of structure formation.  There are two
effects to consider here: fluctuation behavior in the transparent
regime and after reionization.  Let us begin
with the first question.  Closely following recombination, the baryons
are released from drag essentially at rest and thereafter
can grow in pressureless linear theory.  The joining conditions
then imply that
${\frac 3/5}$ of the perturbation joins the growing mode [\PeeblesLSS],
yielding present fluctuations of
$\sim {\frac 3/5} C_I D(z=0)/D(z \sim 800) $
where the residual ionization makes the drag epoch $z \sim 800 < z_*$.
The evolution is
again well described by the interpolation function \eqnDragGrow\
so that $\Delta_b(a) = {\cal G}(a,a_t)C_I$.
By this argument, the effective redshift to employ is $z_t \sim
{\frac 3/5} 800 \approx 400-500$.  We take here
$z_{t} \approx 450$.

Now let us consider the effects of reionization at $z_i$. After $z_i$, Compton
drag again prevents the baryon perturbations from growing.  Therefore the final
perturbations will be $\Delta_b(a_0) \approx \Delta_b(a_i) D(a_0)/D(a_d)$.
Joining the transparent and ionized solutions, we obtain
$$
\Delta_b/C_I = {\cases { {\cal G}(a,a_t)
                        & $a<a_g$  \cr
                {\cal G}(a_i,a_t){\cal G}(a,a_d) ,
                        & $a>a_g$ \cr} }
\eqn\eqnGrowLate
$$
which is plotted in Fig.~14.
Since perturbations do not stop growing
immediately after reionization and ionization after
the drag epoch does not affect the perturbations, we take
$a_g = {\rm min}(1.1a_i,a_d)$.
\smallskip
\goodbreak
\noindent{\largeit B. Photon Evolution: The Doppler and Small Scale Effects}
\smallskip
We now need to examine the evolution of photon temperature perturbations
in light of these results for the matter.  As the diffusion length
overtakes the fluctuation, acoustic oscillations in the photons
are washed away (see Fig.~15).  Since the baryon velocity can grow after
$z_d$ so that
$V_b \gg V_\gamma$,
Doppler shifts off moving electrons will regenerate
temperature perturbations.  Yet since $k > k_D$,
 these perturbations will be erased
as the photons travel across several wavelengths of the perturbation and
are rescattered.
 Unlike the acoustic oscillations, photon evolution before {\it last}
 scattering
is inconsequential.
This is true
even in late ionization scenarios.  The large acoustic fluctuations
frozen in at recombination become anisotropies as they free stream
to the reionization epoch where they are damped as $e^{-\tau}$ by
rescattering.  Thereafter, fluctuations are regenerated by the Doppler
effect at last scattering exactly as in the partially ionized case.
Doppler anisotropies therefore can be completely described by the matter
fluctuations at last scattering [\ref\Kaiser\Kaiser].

Moreover, since the optical depth decreases only due to the expansion, last
scattering
will extend for a period of time comparable to $\eta_*$.
The later last scattering is,
the thicker the last scattering surface.
Cancellation between
photons which last scattered off a crest or trough
of the matter perturbation
will severely damp the Doppler effect on scales smaller
than the thickness.
Together, this implies that the
higher the ionization fraction $x_e$, the more severely damped these
anisotropies will be (see Fig.~15).
One must be careful however to avoid overproducing spectral distortions
in the CMB due to scattering off hot reionized electrons.
Under most plausible ionization scenarios, fully ionized open
models are ruled out by the low Compton-$y$ distortion [\Models].

We can analytically
 account for these effects by using the {\it weak} coupling approximation
[\Kaiser] which
treats the photons as diffusing across independently evolving
baryon perturbations.
Moreover, due to the cancellation of the Doppler effect, ordinarily
negligible contributions become significant and must also
be included in this formalism, \eg\ the late
ISW effect [\HS] and second order contributions
[\ref\Efstathiou \EB,\Efstathiou].  The dominant second order
correction, called the Vishniac effect [\ref\Vishniac\ref\HSS
\Vishniac,\HSS]
couples $V_b$ to the spatial dependence of the scattering probability,
\ie\ since $n_e \approx \left< n_e \right> (1+\Delta_b)$, let
$\dot \tau \rightarrow \dot \tau (1+\Delta_b)$
in equation~\eqnGen.
Note that this effect is {\it not} present in the numerical calculation.
Ignoring curvature and taking the ordinary Fourier transform, we obtain
the formal solution for the $k$th mode of the Boltzmann
equation [\HS],\footnote*{Since
the Vishniac effect is not linear, we must consider $k$-mode coupling.
Therefore, in this and the following
sections where power spectra are employed,
we restore the $k$-index of the perturbations to avoid confusion.}
$$
\left[\Theta + \Psi \right](\eta,k,\mu) = \left[\Theta+\Psi\right](\eta_d,
k,\mu)e^{ik\mu(\eta_d-\eta)}e^{-\tau(\eta_d,\eta)} +
[\Theta_{DSW}+\Theta_{ISW}+\Theta_V](\eta,k,\mu) ,
\eqn\eqnBoltSoln
$$
with  $k\mu = \bk \cdot \bg$, the optical depth $\tau(\eta_1,
\eta_2)=\int_{\eta_1}^{\eta_2} \dot \tau
d\eta$, $\Theta_{DSW}$ the Doppler and SW contributions,
$\Theta_{ISW}$ the late ISW effect and
$\Theta_V$ the Vishniac effect.
As noted above, scattering rapidly
damps out the contributions from before the drag epoch as
$e^{-\tau}$.
Thus the photon
temperature perturbation is a function of the matter perturbations
alone.
These source terms are explicitly given by
$$
\eqalign{
\Theta_{DSW}(\eta,k,\mu) &= \int_{\eta_d}^\eta (\Theta_0 + \Psi
- i \mu V_b) \,\dot \tau e^{-\tau(\eta',\eta)}\,
e^{ik\mu(\eta'-\eta)}
d\eta', \cr
\Theta_{ISW}(\eta,k,\mu) &= \int_{\eta_d}^\eta 2 \dot \Psi
e^{-\tau(\eta',\eta)} \,
e^{ik\mu(\eta'-\eta)}
d\eta', \cr
\Theta_{V}(\eta,k,\mu) &= \int_{\eta_d}^\eta  -i\sum_{k'}
\mu' V_b(\eta',k')
\Delta_b(|\bk-\bk'|)
\,\dot \tau
e^{-\tau(\eta',\eta)}\,
e^{ik\mu(\eta'-\eta)}
d\eta', \cr}
\eqn\eqnSSFISW
$$
where recall that the plane wave decomposition is
defined such that $\bg \cdot \bv_b(\eta,\bx) = -i \mu V_b(\eta,k) \exp(i\bk
\cdot \bx)$ [see equation~\eqnLDecomposition].  The visibility
function $\dot \tau e^{-\tau}$ picks out the epoch of last scattering,
and
the second order nature of the Vishniac term is reflected in the mode
coupling sum.

For scales smaller than the thickness of the last scattering
surface, as determined by the width of the visibility function
 $\dot \tau e^{-\tau}$,
the
Doppler, SW, and Vishniac effects, will be cancelled by oscillation in the
integrand of equation~\eqnSSFISW\ for all but the  perpendicular
$\mu=0$ mode.  Analogously $\dot \Psi e^{-\tau}$ defines a
thickness of the ``gravitational last scattering surface'' under which
contributions are also cancelled.

Linear theory flows are irrotational, $\bg \cdot \bv
\propto \mu k$, and
gravitational redshifts are absent in the direction perpendicular
to the oscillation.
Both
the Doppler effect and the SW
effect thus vanish for $\mu=0$, implying severe cancellation.
Because cancellation occurs similarly and $|\Psi| \ll |V_b|$
on small scales, the residual Sachs-Wolfe effect will be much smaller
that the other two effects.
By angularly
averaging the first of equations \eqnBoltSoln, we obtain the residual
effect on the monopole [\Kaiser],
$$
\Theta_0 \approx {1 \over 4} \Delta_\gamma
\approx {V_b \over k}\dot \tau,
\eqn\eqnMonop
$$
which
feeds back through equation~\eqnBoltSoln\ into
the uncanceled $\mu=0$ mode
[\HSS].  This $\mu=0$ mode is also how all effects avoid
cancellation.
Small scale density perturbations,
with oscillations {\it perpendicular} to the line of sight, can be in bulk
motion {\it parallel} to the line of sight.
The result is the Vishniac effect:
a small scale temperature anisotropy due to the increase in
probability of scattering off an overdense region.
The late ISW effect is similar. But note that in the $\Lambda$
case, the thickness is comparable to $\eta_0$ implying
cancellation occurs up to scales near the present
horizon.  Thus whereas in the open case one sees a gentle
roll off of contributions in $\ell$, in the $\Lambda$ case,
anisotropies fall sharply even from the lowest $\ell$.

This method is valid for calculating these secondary
fluctuations for {\it either} the isocurvature or adiabatic
scenario, under any ionization history in which last scattering occurs
after the end of the drag epoch.\footnote*{For baryonic compact
object dominated models, the density of free electrons may be so
depleted  that last scattering occurs before the drag epoch even
with maximal ionization. Though the analysis is more complicated,
it
remains true that scattering attempts
to equalize
$V_b$ and $V_\gamma$.  This boosts $V_b$ and
suppresses $V_\gamma$ at large scales and vice versa at small
scales [\Models].}
In Fig.~10, we show that
this approximation \eqnMonop\ compares well with the numerical
result
roughly between the drag epoch and last scattering, as
expected.

Integration of equation~\eqnSSFISW\ determines the present rms
temperature perturbations as a function of the underlying matter
fluctuations $P(k)=|\Delta_T(\eta_0,k)|^2$.
For late last scattering, the
integrands in equation \eqnSSFISW\ are wide bell shaped functions.
The functions $\Theta_{DSW}$ and
$\Theta_{ISW}$ are therefore approximately Fourier transforms whose
contribution to the rms can be approximated using Parseval's theorem
[\Efstathiou, \HSS, \Kaiser]
$$
|\Theta + \Psi|_{rms}^2 (\eta_0,k)
= \pi {P(k) \over (k\eta_0)^5}
\int_0^1 dx \left[
|G_{DSW}(x) + G_{ISW}(x)|^2 + |G_V(x)|^2 I_V(k) \right],
\eqn\eqnRpower
$$
where $x=\eta/\eta_0$.
The growth is accounted for by the conformal time integrals over
$$
\eqalign{
G_{DSW}(x) &=  \eta_0^3 \left[ {\ddot D \over D_0} \dot
\tau + {\dot D \over D_0}\ddot \tau \right] e^{-\tau(\eta,\eta_0)}, \cr
G_{ISW}(x) &= 3 {\left( a_0 \over a \right)^2} \eta_0^3
H_0^2 \Omega_0 \left[ {\dot D \over D_0}{a \over a_0} - {D \over D_0}
{\dot a \over a_0}\right] e^{-\tau(\eta,\eta_0)}, \cr
G_V(x) &= {\dot D \over D_0}{D \over D_0} \dot \tau \eta_0^2
e^{-\tau(\eta,\eta_0)}, \cr }
\eqn\eqnGFact
$$
where $D_0 = D(\eta_0)$
and the time independent mode coupling for the Vishniac effect is
[\Efstathiou]
$$
I_V(k) = {V \over 16 \pi^2} {(k\eta_0)^5 \over \eta_0^3}
	\int_0^\infty dy \int_{-1}^{1} d(\cos\theta)
	{(1-\cos^2\theta)(1-2y\cos\theta)^2 \over (1+y^2-2y\cos
	\theta)^2} {P[k(1+y^2-2y\cos\theta)^{1/2}] \over P(k)}
	{P(ky) \over P(k)}.
$$
The Vishniac effect
peaks strongly to small scales, whereas the
first order Doppler and integrated Sachs-Wolfe contribution have
the {\it same} scale dependence reflecting the cancellation process [\HS].

In summary, cancellation occurs because in the diffusive and free
streaming limit photons travel through many wavelengths of
the matter fluctuation source.  Cancellation
is particularly severe for the Doppler and SW effects due to a lack
of a perpendicular mode, but is also present for the late ISW and Vishniac
effects.
\bigskip
\goodbreak
\centerline{\largebf VI. Matter \& Temperature Power Spectra}
\smallskip
The relative
amplitudes of the $k$-modes which form the power spectrum are
often set by
an {\it ab initio} anzatz.  Taking a more agnostic approach which
allows for future empirical determination of the weights,
we examine the transfer functions, \eg\
$P(k) \equiv |\Delta_T(a_0,k)|^2 = |T(k) C_I(k)|^2$ and
$|T(k)C_A(k)|^2$  for isocurvature and adiabatic matter
perturbations
respectively.  For CMB anisotropies, each $\ell$-mode evolves
differently and thus possesses its own
transfer function.
We present here the full $\ell-k$ space structure of the
{\it anisotropy} transfer
function.  To illustrate the effects of altering
the $k$-weighting,
we also present a few specific examples for the underlying
spectrum.  Perhaps the
simplest possible
choice is
a random phase pure power law in
$\tilde k$ initially, \ie\ $|C_A|^2 \propto \tilde k^n$
and $|C_I|^2 \propto \tilde k^n$ for adiabatic and isocurvature modes
respectively.  Although this may not be realistic near
the curvature scale where geometric effects can
introduce novel features [\ref\Lyth \Lyth],
these toy models do illuminate the general case.
\smallskip
\goodbreak
{\noindent \largeit A. Matter Transfer Function}
\smallskip
{\noindent \largeit 1. Adiabatic Models}
\smallskip
For adiabatic models, the matter transfer function is affected by
the dynamics and matter content only.  Since in low $\Omega_0$ models,
matter-radiation equality occurs late, the scale at which
the transfer function turns over due to radiation growth suppression
is larger.  Furthermore, growth in the matter dominated epoch
is suppressed due to curvature and/or $\Lambda$.  Combining
the standard fitting formula for the numerical results
[\ref\Peacock\Peacock]
 with our
analysis of growth rates, we
may write the total transfer function
as
$$
T(k) = D(\eta_0) {\ln(1+2.34q) \over 2.34q}
[1 + 3.89q + (14.1q)^2 + (5.46q)^3 + (6.71q)^4]^{-1/4}
\eqn\eqnTransfer
$$
where $q \equiv k/[\Omega_0 h^2 \exp(-2\Omega_b)]$ and is
valid for $\Omega_b \ll \Omega_0$.    Aside from the small $\Omega_b$
dependence to account for coupling with the
photons,
$q \propto k/k_{eq}$.
For scales that
enter before equality, the perturbations grow as $a^2$ until Jeans
crossing at $a_J \propto (k_{eq}/k)$.  Thus the transfer function
is flat at large scales and goes smoothly to $k^{-2}$ at small scales.

The definition of the adiabatic transfer funtion employed here carries
information about the growth from {\it equality}
to the present in the
form of $D(\eta_0)$.  However, in comparing different $\Omega_0$
models, the epoch of equality shifts.  A more useful choice requires
equal potentials $\Phi$ at the initial epoch
(below
the curvature scale).
Specifically, this amounts to considering the quantity $T(k)/
(\Omega_0 h^2)^2$ due to the $k_{eq}^2$ from the Poisson equation
(see Fig.~16).

This has
the added benefit that the $k$-space SW contribution will also
be the same.  If the total ISW contributions after last scattering are
negligible, this normalization of the transfer function is
identical to a large angle anisotropy normalization for
scale invariant spectra.  For tilted spectra, one must account
for changes in the $k$ to $\ell$ space projection through
$\eta_0-\eta_*$.
Since $n=1$ $\Omega_\Lambda \simlt 0.9$ models approximately satisfy
these conditions at the largest scales (see \S IIID),
the relative amplitude of {\it anisotropy} normalized
matter fluctuations on various scales
can be read directly from Fig.~16.  Matter fluctuations at the
$8 h^{-1}$Mpc ($k \approx 0.1 h $Mpc$^{-1}$)
scale decrease in amplitude for fixed $h$ due to
a change in equality rather than $\Lambda$ growth suppression
[\ref\EBW\EBW].

For open adiabatic models, the situation is more complicated.
As we have shown, the late ISW not the SW effect
dominates the large angle anisotropies.  From this we would
expect that the anisotropy normalized matter amplitude would decrease
relative to Fig.~16.  However, this can be countered by the
suppression of the gravitational potential (SW and ISW)
effects from the curvature term in
the  Poisson equation \eqnPotential.   If the underlying spectrum
is taken to be $|C_A|^2 \propto \tilde k$, these effects in fact
nearly cancel.   However, we have reason to believe that the presence
of a curvature scale may influence the initial conditions.  For
example, in the specific open inflationary case calculated in
[\Lyth], the boost from the late ISW effect dominates and further suppresses
matter fluctuations.

\smallskip
\goodbreak
{\noindent \largeit 2. Isocurvature Models}
\smallskip

In contrast to the adiabatic case, the isocurvature matter transfer function
exhibits relatively complicated structure.
On scales larger than the Jeans length,
the matter gains a
$k^2 - 3K$ tail through the feedback mechanism (see \S IIIB)
and grow as $D(a)$ after radiation domination.
Below this scale, the perturbations have damped
oscillations around the initial conditions $C_I$ until the end of the drag
epoch.
Since the Jeans scale goes to
a constant in the matter dominated epoch, this implies that the transfer
function will have a significant peak at the maximal Jeans scale.
Note that if the universe is not sufficiently matter dominated at
last scattering, this scale could be less than its absolute maximum.
Thus as the ionization fraction decreases, the peak in
the transfer function moves to smaller scales (see Fig.~17).
Since isocurvature models are motivated by the desire to satisfy
observational estimates of $\Omega_0 \approx 0.2$, we will
concentrate on the effects of ionization history rather than
matter content.

We can in fact deduce some of the properties of the oscillatory
regime from our simple analysis.  In \S IVB,
we have shown that the oscillations
decrease as $(1+R)^{-1/4} {k_{eq}/k}$ until they absolutely
disappear for scales smaller than the diffusion length $k \gg k_D$,
leaving a constant tail in the transfer function.
However, after the drag epoch, all
scales grow as $D(a)$ so that the flat tail will have an amplitude
which is dependent on the ionization history.  Notice furthermore
that
the oscillations become less prominent if last scattering is delayed,
since both the $(1+R)^{-1/4}$ and diffusion suppression increases.

For $\Lambda$ models, the change in the growth rate boosts
the amplitude of the transfer function.  Since neither the
Jeans scale nor the drag epoch depends on $\Lambda$ itself, the
shape of the transfer function is the same aside from the lack
of curvature effects at the largest scales.  Analytic
fitting formula may be adapted from the fully ionized case
given by [\ref\CSS\CSS] modified to account for the growth rates
presented in \S VA.

Since large scale structure measurements indicate that $P(k) \propto
k^{-1}$ at intermediate scales
$10^{-2} \simlt k/h \simlt 1$Mpc$^{-1}$,
which fall just below the maximal Jeans scale, the isocurvature
scenario must have an
initial spectrum of $n \approx -1$ at least at
these scales [\ref\SugSuto\SugSuto].
If the initial spectrum is assumed to be a single power
law, this implies a very steep matter power spectrum
at large scales since $\Delta_T \propto (k^2 - 3K) C_I$.  In other
words, an isocurvature spectrum with index $n$ corresponds approximately
to an adiabatic spectrum of $n+4$ at large scales, \eg\
$n = -3$ yields scale invariance.  The steep
$ n = -1$ implies
large amounts of small scale power which
allows for the early collapse of structure and early reionization
[\PeebPIB].
\smallskip
\goodbreak
\noindent{\largeit B. CMB Anisotropies}
\smallskip
{\noindent \largeit 1. Large Angles}
\smallskip
As we have shown in \S IIID, the total Sachs-Wolfe effect can lead
to interesting structure in the anisotropy at large scales in
an $\Omega_0 < 1$ universe.  But how dependent are the
features on the underlying power spectrum?
In Fig.~18, we show the full contributions to the anisotropy in
both $\ell$ and $k$ as given by equation~\eqnWeight\
for adiabatic models.  Although we have chosen to represent
a $|C_A|^2 \propto
\tilde k$ weighting, {\it any} initial power spectrum can be obtained
from scaling by $|C_A|^2/\tilde k$.
The full information
of the two dimensional radiation transfer function is contained here.
Notice that integration in $\log k$
yields the total anisotropy $\propto (2\ell+1)C_\ell$ and in $\log\ell$
gives the rms temperature
fluctuation for a given $k$-mode.

The adiabatic $\Omega_0=1$ case shown to full scale in the top left panel
of Fig.~18, shows the tight $k-\ell$ correlation of the projection from
last scattering (see \S IIID). The
SW effect contributes at large physical scales and the
acoustic peaks at small physical scales.  An expanded view in top
right panel shows the break between the two effects around $\log\ell=1.5$
and $\log(k*$Mpc$) = -2$ for this model.
Pivoting the underlying power spectrum around this
value of $k$ simply emphasizes one effect over the other.

The situation is more complicated for $\Lambda$ and open models.
The ISW term contributes to anisotropies
for intermediate values of $k$.  For low $\Omega_0 \approx 0.1-0.3$
the early ISW effect
fills the gap between the SW ridge
and acoustic peaks of the $\Omega_0=1$ model (see \S IIID and IVC).
The main contribution comes directly after
horizon crossing for these intermediate $k$-values and
thus projects onto lower $\ell$-modes than the SW effect.

At still larger
scales, late ISW contributions become important.
For $\Lambda$ models, they lead to
low $\ell$
contributions since most fluctuations have not had time to free
stream to high multipoles and those which have are cancelled.
For intermediate $k$, the  late ISW effect adds in quadrature to
the SW effect.
Yet for the largest $k$-modes, the SW
effect itself has not free streamed and the late ISW effect
will partially cancel it.  Thus, depending on the $k$-weighting
of the initial power spectrum, the late ISW term can
have different effects.  In Fig.~19, we plot the anisotropies for
single power law weightings.  Notice that the boost in the low multipoles
from $\Lambda$ only occurs for intermediate values of the
slope.
On the other hand, for isocurvature models,
the $\Lambda$ contributions to the
late ISW effect are never prominent due to the dominance of the SW and
early ISW effects.

For $\Omega_0 \approx 0.1-0.3$ open adiabatic
 universes, the total ISW effect
almost always overwhelms the SW effect.
There are two exceptions.  Below a certain scale, the late ISW
effect is thickness cancelled.  Moreover these scales are
often superhorizon sized at radiation domination so that the
early ISW effect does not contribute either. At the largest
scales, the projection carries the late ISW effect onto the
unobservable monopole and dipole.
Thus just as in the $\Lambda$
case, the relative weight of SW versus late ISW increases at large scales.
Again, the SW and late ISW contributions
at the largest scales tend to cancel.  This is more important for
curvature as opposed to $\Lambda$ late ISW contributions
 since the horizon size
at curvature domination is smaller than that at $\Lambda$ domination.
Indeed for somewhat higher $\Omega_0$ open models
($\Omega_0 \approx 0.5 - 0.8$) where the SW and late ISW contributions
are more comparable, cancellation can lead to a suppression of large
angle anisotropies [\SugSilk].

On the other hand, for the largest modes
the amplitude
of late ISW contribution itself
{\it decreases}
with $\ell$ due to the projection.   Yet to have any net
effect, the initial power spectrum must rise sharply to large
scales to counter the $k^2/K$ Poisson equation suppression.
Even the $k^{-1}$ rise toward large scales in recent predictions
of an
open inflationary model
[\Lyth], does not overcome this suppression.
Thus it is difficult to obtain a spectrum with
falling anisotropies in open universe; in most cases the lowest order
multipoles will show a rise in the anisotropy
$\ell$ (see Fig.~19).  This is often followed by a
dip due to the transition between late and early ISW domination.

Open isocurvature models do not suffer Poisson suppression
which makes curvature scale peculiarities
manifest.
Anisotropy contributions come from $k$'s all the way to the
curvature scale $k=\sqrt{-K}$ or $\tilde k = 0$.  Notice that this covers
an infinite range in $\log\tilde k$, and yet the contributions
retain exactly the same $\ell$-space structure.
As we discuss in Appendix A, the radial eigenfunctions $X_\nu^\ell(\chi)$ have
the peculiar property that even as the effective wavenumber $\nu = \tilde
k /\sqrt{-K}\rightarrow
0$, they possess structure on order the curvature scale and are exponentially
suppressed thereafter.
Although the functions are complete, no random phase superposition of
them will ever produce structure above the curvature scale.
As $\tilde k \rightarrow 0$,
all modes contribute at the angle the curvature scale subtends when
the anisotropy was generated,
\eg\ at approximately
the distance
$\eta_0-\eta \approx \eta_0$ for the SW and isocurvature early ISW effect.

For $\tilde k$-scale invariant potential, random phase weighting,
the infinite number
of decades in $\log \tilde k$ as $\tilde k \rightarrow 0$ causes a
divergence in the anisotropy, if no cut off is assumed (see Fig.~20a).
Moreover, {\it any} spectra that places even more power on
scales
$\tilde k \simlt \sqrt{-K}$
will result in the same final anisotropy.
This peculiarity can be seen in Fig.~19 for
open isocurvature models with $n \simlt -3$.
Note however that ``$\tilde k$-scale invariance'' does {\it not}
imply equal power on all {\it physical} scales since all low
$\tilde k$ eigenfunctions have curvature scale power.  Physical
scales above the present horizon do not contribute to
anisotropies despite the apparent divergence from low $\tilde k$.
For the adiabatic case, the suppression of such scales from
the Poisson equation prevents this effect from becoming manifest for
reasonable $n$.

This indicates that for open isocurvature scenarios we must alter the
power spectrum from $\tilde k$-scale invariance to have enough
power at small scales to form galaxies.
For spectra that are strongly tilted to small scales,
anisotropies converge to approximately
$\ell (2\ell+1)C_\ell \propto \ell^2$ and become independent
of $n$ and the model.
This occurs for
$n \simgt 1$ for isocurvature and $n \simgt 5$ for adiabatic conditions where
recall that there is a $k^4$ difference in the correspondence of $n$ to
the matter power spectrum.
Because fluctuations are dominated by the smallest
scale fluctuations present, \ie\ those at the photon diffusion
length $k_D$, equation~\eqnCl\ implies
that $C_\ell$ is constant in $\ell$ as required.
For an isocurvature
scenario with index
$-1 \simlt n \simlt 0$, which is of interest for structure
formation, this asymptotic value has not yet been reached and
$\ell (2\ell+1) C_\ell \propto \ell$ approximately.
This corresponds to an effective COBE DMR slope of
$n_{\rm eff} \approx 2$ [\SugSilk] implying that
isocurvature models have significantly steeper anisotropy
spectrum than the standard
CDM model in which $n_{\rm eff} \approx 1$, but not as steep as one might
naively think. In Fig. 20b, we show such an $n=-1$ weighting.  Notice
that bleeding from smaller $k$-modes than the main $k-\ell$ projection
ridge is responsible for filling in the low $\ell$ anisotropy.

In summary, we have identified several independent
causes of a downturn of anisotropies at low $\ell$:
\smallskip
\noindent\item{(a)}
The Poisson equation curvature cut off.
\smallskip
\noindent\item{(b)}
SW and late ISW cancellation.
\smallskip
\noindent\item{(c)}
Eigenfunction curvature cut off.

\noindent
The first effect only occurs in open adiabatic models and manifests itself
for $\Omega_0 \simlt 0.3$.  The second effect is most significant when
the SW and late ISW effects are comparable, \eg\ open adiabatic models
with $\Omega_0 \approx 0.5 - 0.8$ [\SugSilk] and comes from scales
which are superhorizon sized at the epoch of late ISW generation.
The last effect applies if the initial
spectrum gives significant weight to randomly phased low $\tilde k$
contributions {\it
and} if the contributions are generated early enough to project onto
an anisotropy instead of a monopole fluctuation, \eg\ open isocurvature
models with $n \simlt 3$.

Two effects can give an
upturn relative to the underlying power spectrum
\smallskip
\noindent\item{(a)}
Late ISW contributions.
\smallskip
\noindent\item{(b)}
High $k$-mode power bleeding into low $\ell$.
\smallskip
\noindent The late ISW effect predicts a rise toward
low $\ell$ because of crest-trough cancellation
at small scales.  In a $\Lambda$ universe,
this cutoff scale is on order the present horizon so that contributions
are already falling sharply with $\ell$ at low $\ell$.
For open universes, the late ISW effect contributes earlier
and has a smaller scale cutoff.  Thus the signature of the
anisotropy transfer function is a rise to a plateau
at low $\ell$.  However, since the $k$-modes which contribute to this
effect are the intermediate ones, this effect is only manifest if the
initial spectrum gives them weight.
For pure power laws, this requires a
roughly scale invariant potential: $\tilde k^3 \Phi^2=$ constant.  In
an open universe, the Poisson cutoff can change the plateau to a
dip in the anisotropy at low $\ell$.
For
the opposite case of small scale weighted power spectra, $n \simgt -0.1$
for isocurvature and $n \simgt 3$ for adiabatic, the higher $k$-ridges
in the projection effect contribute strongly to low $\ell$ multipoles.
This implies that there is a maximum slope with which low order multipoles
can rise, $C_\ell\approx$ constant.

If any such features are detected in the observed spectrum
and are statistically significant considering cosmic variance,
some variation of the standard CDM picture will be necessary.
However, even though a simple tilt (single power law)
in the power spectrum
cannot mimic such features, it is clear that more complicated
initial spectra can.
This degeneracy between the initial conditions and the evolutionary
effects is lifted by assuming an {\it ab initio} model.  In this case,
large scale anisotropies are a simple yet powerful probe of the underlying
cosmology as is well known.  Alternatively, once the fundamental
cosmological parameters, \eg\ $\Omega_0$, $h$, $\Lambda$, are
known, they will tell us what the initial conditions for structure
formation are.

\bigskip
\goodbreak
\noindent{ \largeit 2. Intermediate to Small Scale Anisotropies}
\smallskip
In standard recombination scenarios, acoustic oscillations determine
the structure of anisotropies for both adiabatic and isocurvature
modes.  Since these oscillations contain a great
deal of structure, it is obvious that more
 cosmological
and model information can be extracted here than at larger angles.
Moreover,  once coverage of the sky at these angles becomes more complete,
these measurements will be more immune to uncertainties from cosmic
variance.

The angular scale of the
peaks is determined by the projection of the sound horizon at
last scattering onto the sky today and is independent
of the underlying power spectrum. Three
cosmological quantities, with their corresponding dependence on
fundamental parameters, enter into its construction:

\noindent\item(a)
$r_s(\eta_*) \approx f(\Omega_0h^2,\Omega_bh^2)$, the
sound horizon at last scattering,

\noindent\item(b)
$\eta_0-\eta_* = f(\Omega_0 h^2, h, \Omega_\Lambda h^2)$,
the distance to the last scattering surface,

\noindent\item(c)
$K = f(\Omega_0 h^2, h, \Omega_\Lambda h^2$), the curvature.

\noindent The first task is to distinguish between
adiabatic and isocurvature scenarios.
For
adiabatic models, the peak $\ell$ values follow the series $(1:2:3:4...)$,
whereas
for isocurvature models $(1:3:5:7...)$.  Since the first peak is
contaminated by the early ISW effect
(see Fig.~21),
the higher peaks are
the most reliable measure of this effect.  On the other hand, this
rise to the first peak can also be used to separate isocurvature
from adiabatic models.  We have noted in \S IVC that the first
isocurvature oscillation is low in amplitude.  Only in adiabatic
models does the first oscillation truly stand out as a peak.

Once adiabatic and isocurvature models are distinguished,
the location of the peaks is uniquely predicted by the cosmological
parameters.  However, the degeneracy in the dependence on
$\Omega_0 h^2$, $\Omega_\Lambda h^2$, $h$, and $\Omega_b h^2$
does not allow inversion of the relation [\ref\Confusion\Confusion].
For example,
an $\Omega_b=\Omega_0=1$, $h=0.5$ adiabatic
model predicts $\ell \sim 400$
and $h=1.0$, $\ell \sim 500$ which can mimic projection effects
from curvature and $\Lambda$.   Of course, if one is willing to
restrict $\Omega_b h^2$ to lie within the nucleosynthesis bounds,
its effect on $r_s(\eta_*)$ is negligible.
On the other hand, the geodesic deviation due to
$K$ with low $\Omega_0$ and $\Lambda=0$
is a severe and easily tested effect.  If the angular
location of the peaks turn out to be multiples of a
high $\ell \simgt 400-500$, then curvature must almost certainly be present in
the model since no reasonable change in $r_s(\eta_*)$ or $\eta_0-
\eta_*$ can account for it [\ref\KSS\KSS].

Since isocurvature acoustic oscillations are likely to be
erased by reionization, let us concentrate on lifting
the degeneracy for the more
plausible adiabatic case.
We can use the deviation of the first
peak from the acoustic series predicted above for this purpose.
The early ISW effect pushes the peak to
larger scales for low $\Omega_0 h^2$ universes.  Moreover, the
amplitudes of the peaks contain a large amount of cosmological
information as well.  Even though this  depends on
the underlying power spectrum, a minimal
assumption, such as a pure power law only over the range of the
peaks, would be sufficient to allow
interesting constraints on cosmological parameters.
As we have seen,
\smallskip
\noindent\item
(a) Lowering $\Omega_0 h^2$ boosts the first peak relative
to the higher peaks due to early ISW contributions
\smallskip
\noindent\item
(b) Raising $\Omega_b h^2$ boosts the odd numbered peaks over
the even due to reduction in the pressure relative to
the gravitational force.
\smallskip
\noindent
In fact, these opposing $h$ dependences nearly cancel for the
first peak if $\Omega_0=1$ and $\Omega_b h^2$ is given by big bang
nucleosynthesis.  This is not true for the higher peaks [\HSCDM,\Uros].
Thus the {\it relative}
amplitudes of the
series of peaks contain crucial cosmological information.
These important tests will depend on having experimental information
for anisotropies $\ell \simgt 200$.

Considering the present experimental focus on $\ell \simlt 200$
anisotropies, it would be useful to extract information from
the ratio of large to intermediate angle anisotropies.
For instance, in the $n=1$ model of Fig.~21a, the rise to
the first peak is more dramatic in low $\Omega_0h^2$ universes.
Unfortunately,
this of course depends on the specific model in question.
However
in general, lowering $\Omega_0 h^2$ increases the intermediate
anisotropies through the early ISW effect whereas increasing
$\Omega_bh^2$ does the same through the acoustic oscillations.
For large scales, the late ISW effect can boost anisotropies
a comparable amount in the open but not the $\Lambda$ case.
However one must recall that in the open
case there is also Poisson suppression of the power spectrum and
other curvature effects.

Of course, allowing the thermal history to deviate from the
standard recombination scenario introduces another degree of
freedom which complicates the extraction of cosmological
information.  If reionization is low, the acoustic peaks
which are damped as $e^{-\tau}$ below the horizon at
last scattering,  may still be observable.  However if
the ionization is high, the detailed information in the
acoustic oscillations is lost to us.  This is likely to
be the case for isocurvature models.  If
the initial spectrum is chosen to be consistent with large scale
structure $n \approx -1$, the large fluctuations at small
scales could result in reionization.
Normalized to
large scale anisotropies, standard recombination
models also produce excessively
large intermediate scale
adiabatic oscillations in the standard recombination scenario
(see Fig.~21b).  Reionization is therefore also {\it necessary}.

In this case the sole
feature is the damping scale which measures the photon diffusion
length at last scattering.
In Fig.~21b, we show the effects of altering the ionization history
of open and $\Lambda$ isocurvature models.
Assuming a cosmological model, the damping scale
fixes the ionization history.  On the other hand, assuming an
ionization history (\eg\ fully ionized), it essentially probes
the horizon size at last scattering as projected via geodesic
deviation.
Although $\Lambda$
models are older and yield a larger distance to the last scattering
surface, the geodesic deviation effect pushes the damping scale
of open models to even smaller angles.  Notice that this also makes
the open universe large angle anisotropies nearly independent of
ionization history since these angles correspond to superhorizon
scales at last scattering.

As for the amplitude of the regenerated fluctuations, we may employ
the analysis of \S VB, to gain insight into the numerical results.
In Fig.~22, we show a comparison
of isocurvature temperature power spectra from the numerical and analytical
calculations.  The numerical calculations are purely first order
and do not include the Vishniac contribution.
The Doppler and SW (DSW)
fluctuations are increasingly suppressed by thickness cancellation as
last scattering is delayed, as reflected in the time integrals of equation
\eqnGFact.
The late ISW effect is of
course independent of ionization but increases as $\Omega_0$ decreases.
For the fully ionized, low $\Omega_0=0.1$ universe shown here,
the late ISW contribution thus
more than doubles the temperature fluctuations at intermediate scales.
On the other hand, the Vishniac effect
depends quadratically on the amplitude of the matter fluctuations
and thus is larger for later last scattering.  In our detailed
numerical study [\Models], we show how these various effects can be combined
to yield the minimal anisotropies for the isocurvature model.

Reionized adiabatic models look similar to isocurvature models
in that the sole feature is at the diffusion scale
at last scattering.   If no underlying power spectrum is
assumed, it may be difficult to distinguish between the
two.
However, as large scale structure measurements reach to larger scales
and CMB experiments to smaller scales, it will be possible to
entirely remove the ambiguity of the initial power spectrum
(see \eg\ [\ref\TegTegTeg\TegTegTeg]).
Consistency between the matter and radiation power spectrum is
indeed the ultimate test of any model for structure formation.  As
we have seen, the difference in the
matter and temperature transfer functions on the {\it same} scale
can remove all doubt on the question of
adiabatic vs. isocurvature initial conditions and/or
standard recombination vs. reionized thermal histories.
\bigskip
\goodbreak

\bigskip
\centerline{\largebf VI. Discussion}
\smallskip
We have comprehensively studied the evolution of density and
temperature perturbations with an arbitrary spectrum of
 adiabatic and isocurvature perturbations
in a critical $\Omega_0=1$, open, and $\Lambda$
dominated expanding universe.  By employing an analytic treatment,
we provide model independent insight into the formation of anisotropies
that is confirmed by its agreement with the full numerical
calculation.
It thus becomes
possible to separate and interpret
each physical process that generates these
perturbations.

Our treatment identifies numerous sources of anisotropies.
Curvature effects due to geodesic deviation and on the fluctuations
themselves give rise to peculiarities in the anisotropy spectrum which
may soon be constrained by the observations.
Moreover gravitational redshift effects due to
the photon's climb out of the potential well (SW effect) as well as
decay or growth in the potential due to radiation (early ISW effect) and
the decay due to the
rapid expansion in an open or $\Lambda$ dominated universe (late ISW effect)
carry specific signatures that may be identifiable in the large
angle anisotropies.  However, the manifestation of these effects
in a particular model will depend on the initial power spectrum.
In examining the dependence on initial conditions, we also present
a particularly simple derivation of the ${1 \over 3}$ (adiabatic) and $2$
(isocurvature)
coefficients multiplying the gravitational potential in the SW
effect.

Smaller angle anisotropies carry information which is less dependent
on the power spectrum.  We have investigated the nature of acoustic
oscillations which give rise to peaks in the anisotropy as well as
diffusion damping which is responsible for its small angle cutoff.
Moreover, we have provided a very simple formula which predicts the
angular location of the peaks as a function of the matter content
and geometry of the universe.  The physical origin of their relative
heights is also clarified.
In reionized models however, acoustic oscillations are damped and give way
to last scattering effects due to baryons in infall.
At intermediate scales, this leads to the Doppler effect whereas
at small scales significant second order Vishniac contributions
must be considered.

Although the principles outlined here are valid for any model,
they can also be used to evaluate {\it currently} popular models
for structure formation.
At the present however, it is not even clear which model, if any, is
consistent with the
large scale structure data alone, much less the detailed
features in the CMB anisotropies.
Despite the success of the elegantly simple
 standard CDM model for
structure formation, it is becoming increasingly clear that
{\it some} modification either in the model or our understanding
of its implications is necessary (\eg\ see [\ref\Ostriker\Ostriker]
for a review).  Normalized to large scale
anisotropies, standard CDM predicts matter fluctuations which imply
a moderately anti-biased picture of galaxy formation [\ref\Bunn\Bunn]
and more
small scale power than is observed for peculiar velocities.  It is
also difficult to understand the dynamical measurements of
a low $\Omega_0$ at small scales in this picture [\ref\Dekel\Dekel].
The obvious solutions within
the context of CDM are to either change the initial power spectrum
from Harrison-Zel'dovich $n=1$, or lower $\Omega_0$ to move the
equality cut off to larger scales.  Indeed the shape of
the matter power spectrum alone seems to indicate $\Omega_0 h \approx
0.25$ [\Peacock], and determinations
of a high Hubble constant $h \approx
0.7-0.8$, if confirmed,
 also support low $\Omega_0$ models due to the age problem
[\ref\Jacoby\Jacoby].

We have fully examined the consequences for anisotropies of these
standard solutions.  The signature of low $\Omega_0$ models at
large scales depends on the underlying power spectrum.  Particularly
in the case of open models, where we {\it expect} deviations from a
single power law spectrum at the curvature scale, this ambiguity
can change the relative amplitudes of anisotropies to matter fluctuations
as well as the shape of the large scale anisotropies
themselves.  For $\Lambda$ models, this is perhaps less of a
concern.  The boost in low order multipoles from the late ISW
effect can be used to constrain $n=1$ models [\ref\BunnSugiyama
\BunnSugiyama].
The acoustic peaks provide a better handle on
the underlying cosmology from both their angular location and
{\it relative} heights.  Even with complications such as
gravitational wave contributions,
which can boost the large scale anisotropy
relative to the matter [\ref\GravWave\GravWave],
 the information contained in the acoustic
peaks is not lost.

Another possible alternative is to abandon adiabatic fluctuations
in favor of isocurvature ones.  This model also changes
the relative amplitude of matter versus temperature perturbations.
However given the likelihood of reionization, the thermal history
of baryonic isocurvature
models can be adjusted to match the observations.  The fundamental
probe here is the slope of the matter and
temperature power spectra.  Present indications are that
$n \approx -1$ ($n_{\rm eff}=2$)
from large scale structure measurements.
The implied discrepancy with flat
large scale anisotropies with $n_{\rm eff} \approx 1$
[\ref\Gorski\Gorski] is beginning to indicate that
no single power law model is adequate [\Hu].  While this
is not necessarily surprising for the open version, it would
require a dramatic break in the power spectrum to counter the
heavily small scale weighted power required by large scale
structure.
Perhaps more damaging to this model is the growing body of intermediate
scale $\ell \approx 50-200$ anisotropy measurements.
  If a steep rise toward $\ell \approx 200$ is
also confirmed [\ref\ScottWhite\ScottWhite], there will also have to be
an additional break below the curvature scale.
Furthermore, there are indications that even large
scale structure measurements themselves do not fit with
single initial power law isocurvature models due to features in
the matter transfer function [\Peacock].

Finally a change in the matter content, \eg\ adding massive neutrinos
[\ref\MDM\MDM] or topological defects [\ref\Top\Top], is
another possibility.  Although we do not explicitly consider
such exotic models, the principles outlined here remain valid.
Sachs-Wolfe contributions
and acoustic oscillations are determined from
the gravitational potential in the same way in these models.
Thus once the evolution of the matter is understood, the
implications for anisotropies is apparent.

Given that none of these alternatives provide a compelling
{\it ab initio} model for structure formation, it is perhaps
best to keep an open mind to all of these possibilities.
As the large scale structure and CMB anisotropy
 data continue to accumulate, the general principles
formulated here will aid in the empirical reconstruction of
a consistent model for structure formation.
\bigskip
\goodbreak
{\it\baselineskip =12pt
\qquad\qquad Many are those under
heaven who attend to their theories and
techniques,

\qquad\qquad and they all believe that nothing can be
added to the ones they possess.

\qquad\qquad Where is the true way
of old to be found?
\bigskip
\hskip 4.5truecm In a haze! Where am I going?

\hskip 4.5truecm In a daze! Where shall I arrive?

\hskip 4.5truecm With the myriad things before me,

\hskip 4.5truecm None will serve as final destination.

\smallskip
\hskip 6.5truecm --Chuang-tzu}
\bigskip
\centerline{\largebf Acknowledgements}
\smallskip
We would like to thank D. Scott, J. Silk, M. White,
and anyone with the patience to read this far!
W.H. acknowledges support from the NSF and N.S. from a JSPS fellowship.

\smallskip
\eject
\bigskip
\goodbreak

\centerline{\largebf References}
{\baselineskip=16pt
\bigskip
\refs[\WSS] M. White, D. Scott, and J. Silk, Ann. Rev. Astron.
Astrophys., {\bf 32}, 319 (1994).

\refs[\PeebPIB] P.J.E. Peebles, Astrophys. J. Lett., {\bf 315}, L73
(1987); P.J.E. Peebles, Nature, {\bf 327}, 210 (1987).

\refs[\EB]  G. Efstathiou and J.R. Bond, Mon. Not.
Roy. Astron. Soc., {\bf 227}, 33p (1987).

\refs[\Models] W. Hu and N. Sugiyama, Astrophys. J., (in press).

\refs[\SugSilk] N. Sugiyama and J. Silk, Phys. Rev. Lett, {\bf 73},
509 (1994).

\refs[\HSCDM] W. Hu and N. Sugiyama, Astrophys. J., (submitted 1994).

\refs[\DSZ] A.G. Doroshkevich, Ya. B. Zel'dovich, R.A. Sunyaev,
Sov. Astron, {\bf 22}, 523 (1978).

\refs[\KS] H. Kodama and M. Sasaki, Int. J. Mod. Phys., {\bf
A1}, 265 (1986).

\refs[\SW] R.K. Sachs and A.M. Wolfe, Astrophys. J., {\bf 162},
815 (1970).

\refs[\Wilson] M.L. Wilson, Astrophys. J., {\bf 273}, 2 (1983).

\refs[\BEMNRAS] J.R. Bond and G. Efstathiou, Mon. Not. Roy.
Astron. Soc., {\bf 226}, 665 (1987); J.R. Bond, in The Early
Universe, eds. W.G. Unruh and G.W. Semenoff, (Dordrecht, Boston)
p. 283.

\refs[\SilkDamp] J. Silk, Astrophys. J., {\bf 151}, 459 (1968).

\refs[\SZDop] R.A. Sunyaev and Ya. B. Zel'dovich, Astrophys. Sp.
Sci., {\bf 7}, 3 (1970).

\refs[\Bardeen] J.M. Bardeen,  Phys. Rev., {\bf D22}, 1882 (1980).
Note his $\Phi_H=\Phi$ and $\Phi_A=\Psi$.

\refs[\KSPert] H. Kodama and  M. Sasaki,  Prog. Theor. Phys. Suppl.,
{\bf 78}, 1 (1984).

\refs[\Mukhanov] V.F. Mukhanov, H.A. Feldman, and R.H
Brandenberger, Phys. Rep., {\bf 215}, 203 (1992).

\refs[\GSS] N. Gouda, M. Sasaki, Y. Suto, Astrophys. J.,
{\bf 341}, 557 (1989).

\refs[\Liftshitz] E.M. Liftshitz and I.M. Khalatnikov, Adv.
Phys., {\bf 12}, 185 (1963).

\refs[\Harrison] E. R. Harrison, Phys. Rev., {\bf D1}, 2726
(1970).

\refs[\AbSch] L.F. Abbott and R.K. Schaefer, Astrophys. J.,
{\bf 308}, 546 (1986). Their definition of the radial
eigenfunctions is equivalent to our $M_\ell^{1/2} X_\nu^\ell$
(see also [\NoteStable]).

\refs[\Precursor] N. Gouda, N. Sugiyama, and M. Sasaki, Prog.
Theor. Phys., {\bf 85}, 1023 (1991).

\refs[\NoteStable] The stability problem can be avoided by two
tricks: rewrite
the Boltzmann equation with $\Delta_\gamma$ replacing $\Theta_0$ and
$\Theta'_\ell = M_\ell^{1/2} \Theta_\ell$ instead of $\Theta_\ell$.

\refs[\Gnedin] N.Y. Gnedin and J.P. Ostriker, Astrophys. J.,
{\bf 400}, 1 (1992).

\refs[\COP] R. Cen, J.P. Ostriker, and P.J.E. Peebles, Astrophys. J.,
{\bf 415}, 423 (1993).

\refs[\SG] N. Sugiyama and N. Gouda, Prog. Theor. Phys.,
{\bf 88}, 803 (1992).

\refs[\Hu] W. Hu, in CWRU CMB Workshop: 2 Years after COBE,
eds. L. Krauss \& P. Kernan, (World Scientific, Singapore), in press.

\refs[\HS] W. Hu, N. Sugiyama, Phys. Rev., {\bf D50}, 627 (1994).

\refs[\Kofman] L. Kofman and A. Starobinskii, Sov. Astr. Lett,
{\bf 11}, 271 (1985).

\refs[\KamSper] M. Kamionkowski and D. Spergel, Astrophys. J.,
{\bf 432}, 7 (1994).



\refs[\Jorgensen] H.E. J{\o}rgensen, E. Kotok, P. Naselsky, and
I. Novikov, Astron. Astrophys.,
(in press).

\refs[\Fernando] F. Atrio-Barandela, and A.G. Doroshkevich, Astrophys. J.,
{\bf 420}, 26 (1994).

\refs[\Uros] U. Seljak, Astrophys. J., (submitted 1994).

\refs[\PY] P.J.E. Peebles and J.T. Yu, Astrophys. J., {\bf 162},
85 (1970).

\refs[\PeeblesLSS] P.J.E. Peebles, Large Scale Structure of the
Universe, (Princeton University, Princeton 1980).

\refs[\SZ] R.A. Sunyaev and Ya. B. Zel'dovich, Astrophys. Sp. Sci.,
{\bf 9}, 368 (1970).

\refs[\HSSy] W. Hu, D. Scott, and J. Silk, Astrophys. J. Lett., {\bf 430},
L5, (1994).

\refs[\JW] B.J.T. Jones and R.F.G. Wyse, Astron. Astrophys.,
{\bf 149}, 144 (1985).

\refs[\Kaiser] N. Kaiser, Astrophys. J., {\bf 282}, 374 (1984).

\refs[\Efstathiou] G. Efstathiou, Large Scale Motions in the
Universe: A Vatican Study Week, eds. Rubin, V.C. and Coyne, G.V.,
(Princeton University, Princeton, 1988) pg. 299.

\refs[\Vishniac] J.P. Ostriker and E.T. Vishniac, Astrophys. J.
{\bf 306}, 51 (1986); E.T. Vishniac, Astrophys. J., {\bf 322}, 597
(1987).

\refs[\HSS] W. Hu, D. Scott, and J. Silk, Phys. Rev.,
{\bf D49}, 648 (1994).

\refs[\Lyth] D.H. Lyth and E.D. Stewart, Phys. Lett.,  {\bf B252},
336 (1990); B. Ratra and P.J.E. Peebles, Astrophys. J. Lett., {\bf 432}, L5
(1994).

\refs[\Peacock] J.A. Peacock and S.J.
Dodds, Mon. Not. Roy. Astron. Soc., {\bf 267}, 1020 (1994).

\refs[\EBW] G. Efstathiou, J.R. Bond, and S.D.M. White,
Mon. Not. Roy. Astron. Soc., {\bf 258}, P1 (1992).

\refs[\CSS] T. Chiba, N. Sugiyama, Y. Suto, Astrophys. J.,
{\bf 429}, 427 (1994).

\refs[\SugSuto] T. Suginohara and Y. Suto, Astrophys. J., {\bf 387}, 431
(1992).

\refs[\Confusion] J.R. Bond, \etal, Phys. Rev. Lett, {\bf 72}, 13,
1994.

\refs[\KSS] M. Kamionkowski, D.N. Spergel, and N. Sugiyama,
Astrophys. J. Lett., {\bf 426}, L57 (1994)

\refs[\TegTegTeg] K. Gorski, Astrophys. J. Lett., {\bf 370}, L5 (1989);
M. Tegmark, E. Bunn, and W. Hu, Astrophys. J., {\bf 434},
1 (1994).

\refs[\Ostriker] J.P. Ostriker, Ann. Rev. Astron. Astrophys.,
{\bf 31}, 689 (1993).

\refs[\Bunn] E. Bunn, D. Scott, and M. White, Astrophys. J. Lett, submitted
(1994).

\refs[\Dekel] A. Dekel, \etal, Astrophys. J., {\bf 412}, 1 (1993).

\refs[\Jacoby] G. Jacoby, \etal, PASP, {\bf 104}, 599 (1992).

\refs[\BunnSugiyama] E. Bunn and N. Sugiyama, Astrophys. J. Lett,
submitted (1994).

\refs[\GravWave] M.S. Turner, M. White, J.E. Lidsey, Phys. Rev.,
{\bf D48}, 4613 (1993); R. Crittenden \etal, Phys. Rev. Lett.,
{\bf 71}, 324 (1993).

\refs[\Gorski] K. Gorski, \etal, Astrophys. J. Lett., {\bf 430}, L89 (1994)

\refs[\ScottWhite] D. Scott and M. White,
in CWRU CMB Workshop: 2 Years after COBE,
eds. L. Krauss \& P. Kernan, (World Scientific, Singapore), in press.

\refs[\MDM] M. Davis, F.J. Summers, D. Schlegel, Nature, {\bf 359},
393 (1992); A. Klypin, J. Holtzman, J. Primack, E. Regos, Astrophys.
J.,
{\bf 416}, 1 (1993)

\refs[\Top] U.-L. Pen, D.N. Spergel, and N. Turok, Phys. Rev.,
{\bf D49}, 692 (1994).


}
\eject
\centerline{\largebf Appendix A: Open Universe Normal Modes}
\bigskip
\noindent{\largeit 1. The Radial Representation}
\smallskip
Fluctuations in an open universe must be decomposed in
the eigenfunctions of the Laplacian
$\gamma^{ij} Q_{|ij} = -k^2 Q$.  To gain intuition about these
functions, let us examine an explicit representation.
In radial coordinates
the 3-metric becomes
$$
\gamma_{ij} dx^{i} dx^{j} = -K^{-1} [d\chi^2 + \sinh^2\chi
(d\theta^2 + \sin^2\theta d\phi^2)],
\Aeqn\AeqnMetric
$$
where recall $\chi = \sqrt{-K}\eta$.  Curvature makes the
surface area of a shell
at distance $\eta$ increase as $-K^{-1} e^{2\chi}$
rather than $\eta^2$ for super-curvature
distances $\chi \gg 1$.  The Laplacian can now be written as
$$
\gamma^{ij} Q_{|ij} = -K \sinh^{-2} \chi
\left[
      {\partial \over \partial \chi}
\left(
      \sinh^2\chi {\partial Q \over \partial \chi }
\right)
+ \sin^{-1}\theta {\partial \over \partial \theta}
\left(
      \sin\theta  {\partial Q \over \partial \theta}
\right)
+ \sin^{-2}\theta {\partial^2 Q \over \partial \phi^2}
\right].
\Aeqn\AeqnLaplacian
$$
Since the angular part is independent of curvature, we
may separate variables such that
$Q = X_\nu^\ell(\chi)Y_\ell^m(\theta,\phi)$
where
$\nu^2 = \tilde k^2/(-K)= -(k^2/K+1)$.
{}From equation \Adis\AeqnLaplacian, it is obvious that
the spherically
symmetric $\ell =0$ function is
$$
X_\nu^0(\chi) = {\sin(\nu\chi) \over \nu \sinh\chi}
= \sqrt{-K} {\sin(\tilde k \Delta \eta) \over \tilde k
\sinh(\Delta\eta\sqrt{-K})}.
\Aeqn\AeqnXIso
$$
As expected, the change in the area element from a flat to
curved geometry causes $\sqrt{-K}\eta
\rightarrow \sinh\chi$ in the denominator.
The higher modes are explicitly given by
[\Liftshitz, \Harrison]
$$
X_\nu^\ell(\chi) = (-1)^{\ell+1} M_\ell^{-1}
\nu^{-2} (\nu^2+1)^{-\ell/2}
\sinh^\ell \chi
{d^{\ell+1} (\cos \nu \chi) \over d(\cosh \chi)^{\ell+1}},
\Aeqn\AeqnHorribleFunc
$$
and becomes $j_\ell(k\Delta\eta)$ in the flat space limit.
Here
$$
M_\ell(\tilde k)  \equiv {(\tilde k^2 -K)...(\tilde k^2 -K\ell^2)
\over (\tilde k^2 -K)^\ell},
\Aeqn\AeqnMFact
$$
which reduces to unity as $K \rightarrow 0$.  It represents our
convention for the normalization of the open universe functions.

It is often more convenient to generate these functions
from their recursion relations [\AbSch].
One such recursion relation is
$$
{d \over d \eta} X_\nu^\ell = {\ell \over 2\ell+1} k X_\nu^{\ell-1}
+{{\ell+1}\over{2\ell+1}}\left[1 - \ell (\ell+2){K \over k^2}\right]
 k X_\nu^{\ell+1},
\Aeqn\AeqnRecursion
$$
which is of the same form as the Boltzmann equation \eqnHierarchy\
for $(\Theta+\Psi)/(2\ell+1)$
in the free streaming limit.  This is quite natural since free streaming
photons arrive at the observer on radial geodesics as an
examination of equation \Adis\AeqnMetric\ shows.  Thus the
solution of the free streaming Boltzmann equation in the absence
of the ISW term is obvious:
$$
{\Theta_\ell(\eta,\tilde k) \over 2\ell+1} =
[\Theta_0+\Psi](\eta_*,\tilde k) X_\nu^\ell (\chi - \chi_*),
\Aeqn\AeqnSW
$$
where we have assumed that the boundary condition at last scattering
is given by the monopole fluctuation as is appropriate to the SW
effect.
The ISW effect acts like an impulse $(\dot \Psi - \dot
\Phi) \delta \eta$  at some intermediate time $\eta$ which then free
streams to the present.  The full solution therefore is
$$
{\Theta_\ell(\eta,\tilde k) \over 2\ell+1} =
[\Theta_0+\Psi](\eta_*,\tilde k) X_\nu^\ell (\chi -\chi_*)
+ \int_{\eta_*}^{\eta} [\dot \Psi - \dot \Phi](\eta',\tilde k)
X_\nu^\ell (\chi-\chi') d\eta'.
\Aeqn\AeqnISW
$$

Let us now examine the peculiar nature of the eigenfunctions.
Since they are complete for
$k \ge \sqrt{-K}$, \ie\ $\tilde k \ge 0$,
should $2\pi/k$ or $2\pi/\tilde k$ be considered the
effective wavelength? In Fig.~23, we plot the spherically
symmetric $\ell=0$ mode given by equation~\Adis\AeqnXIso.
The argument in favor of $\tilde k$ is that
its first zero is at $\Delta \eta = \pi/\tilde k$. This is
related to the completeness property: the zero crossing property
shows that as $\tilde k \rightarrow 0$ we can obtain
arbitrarily large structures.  However even in this limit,
the amplitude of the structure above the curvature scale is
suppressed as $e^{-\chi}$.  The effective scale of the
{\it prominent} structure thus goes to the curvature scale
favoring $k^{-1} = 1/\sqrt{-K}$ as the effective wavelength.
In fact, the $e^{-\chi}$ behavior is {\it independent}
of the wavenumber and $\ell$,
if $\chi \gg 1$.

This peculiarity in the eigenmodes has significant consequences.
Any random phase superposition of the eigenmodes $X_\nu^\ell$
will have exponentially suppressed structure larger than
the curvature radius.  Even though completeness tells us that
arbitrarily large structure can be built out of the $X_\nu^\ell$
functions,  it {\it cannot} be done without correlating the modes.
This is even if the structure has support only to a finite radius
which is above the curvature scale.

Is the random phase hypothesis and the lack of structure
above the curvature scale reasonable?
The fundamental difference between open and flat universes is
that the volume increases exponentially with the radial
coordinate above the curvature scale
$V(\chi_c) \sim [\sinh(2\chi_c) -2\chi_c]$ as the line element
of equation \Adis\AeqnMetric\ shows.
Structure above the curvature scale implies correlations
over vast volumes [\KamSper].
It is in fact difficult to
conceive of a model where correlations do not die exponentially
above the curvature radius.  The random phase hypothesis has
been proven to be valid for {\it adiabatic}
 inflationary perturbations [\Lyth].
However, a definitive answer to this question for {\it isocurvature}
models awaits the invention of a mechanism for generating
such perturbations in a consistent model for structure formation.

\bigskip
\noindent{\largeit 2. General Angular Functions}
\smallskip
Although the radial representation
suffices for many purposes, often one needs the full
machinery of the general normal mode decomposition.
Formally, the angular and
spatial fluctuations of the full radiation field is decomposed
into [\Wilson]
$$
\Theta(\eta,\bx,\bg) =
\sum_{\ell=0}^{\infty} \Theta_\ell(\eta,k)G_\ell (\bx,\bg),
\Aeqn\AeqnLDecomposition
$$
where
$$
G_\ell(\bx,\bg) =
(-k)^{-\ell} Q_{|i_1...i_\ell}(\bx)P_\ell^{i_1...i_\ell}(\bx,\bg),
\Aeqn\AeqnGl
$$
and
$$
\eqalign{
P_{0} & = 1,  \qquad
P^i_{1} = \gamma^i, \cr
P^{ij}_{2} & = {1 \over 2}(3\gamma^i \gamma^j - \gamma^{ij}), \cr
P^{i_1...i_{\ell+1}}_{\ell+1} &=
{2\ell+1 \over \ell+1} \gamma_{\vphantom{\ell}}^{(i_1}
P_\ell^{i_2...i_{\ell+1})}
 - {\ell \over \ell+1} \gamma_{\vphantom{\ell}}^{(i_1
i_2}P_{\ell-1}^{i_3..i_{\ell+1})}, \cr
}
\Aeqn\AeqnP
$$
with parentheses denoting symmetrization about the indices.
For flat space, this becomes
$G_\ell = (-i)^\ell \exp(i\bk \cdot \bx)P_\ell(\bk \cdot \bg)$,
where $P_\ell$ is an ordinary Legendre polynomial.
Notice that along a path defined by fixed $\bg$, the flat
$G_\ell$ becomes
$j_\ell(k\eta)$ after averaging over $k$-directions.
Travelling on a fixed direction away from a point is
the same as following a radial path outwards.  Thus fluctuations {\it along}
this path can be decomposed in the radial eigenfunction.  We shall see
that this argument can be generalized to the open universe case and
allows one to interpret equation~\Adis\AeqnGl\ more easily.

We can also use the properties of $G_\ell$ to simplify the Boltzmann
equation~\eqnGen.
The anisotropic stress perturbation of the photons, defined as
$$
\eqalign{
\Pi^{ij}_{\gamma} &\equiv
4 \int {d\Omega \over 4\pi} \left(\gamma^i \gamma^j
- {1 \over 3} \gamma^{ij} \right) \Theta(\eta,\bx,\bg), \cr}
\Aeqn\AeqnAniso
$$
is therefore related to the quadrupole moment,
$$
{1 \over 16} \gamma_i \gamma_j \Pi^{ij}_\gamma = {1 \over 10}
\Theta_2 G_2.
\Aeqn\AeqnAnisoQuad
$$
The recursion relation
$$
\eqalign{
\gamma^i G_{\ell|i}
& = {d \over d\eta}G[\bx(\eta),\bg(\eta)] =
\dot x^i {\partial \over \partial x^i} G_\ell + {\dot \gamma^i}{\partial \over
\partial \gamma^i}
G_\ell \cr
&= k \left\{ {\ell \over 2\ell+1} \left[1-(\ell^2-1)
{K \over k^2} \right]G_{\ell-1} - {\ell+1 \over 2\ell+1} G_{\ell+1} \right\},
\cr}
\Aeqn\AeqnLLPM
$$
which follows from equation~\Adis\AeqnGl\ and \Adis\AeqnP\
[\Precursor],
completes the simplification of equation~\eqnGen\ to \eqnHierarchy.
Here we take $\bx(\eta)$ to be the integral path along $\bg$.
By comparing equations~\Adis\AeqnRecursion\ and \Adis\AeqnLLPM, the open
universe generalization of the relation between $G_\ell$ and
the radial eigenfunction is now apparent:
$$
G_\ell[\bx(\eta),\bg(\eta)] = M_\ell X_\nu^\ell(\eta).
$$
The only conceptual difference is that for the radial path that we decompose
fluctuations on, $\bg$ is not constant.
This also clarifies the interpretation of the recursion relation for
$G_\ell$ [equation~\Adis\AeqnLLPM].
Finally
by employing these definitions, we may write the temperature correlation
function as [\Wilson]
$$
\left< \Theta^*(\eta_0,\bx,\bg)\Theta(\eta_0,\bx,\bg')\right>
= {V \over 2\pi^2} \int {d\tilde k \over \tilde k}
\sum_\ell {M_\ell(\tilde k)
\over 2\ell+1}
\tilde k^3 |\Theta_\ell(\eta_0,\tilde k)|^2 P_\ell(\bg \cdot \bg'),
\Aeqn\AeqnHowIsItDerived
$$
where $P_\ell$ is a Legendre polynomial.  This implies the
definition of $C_\ell$ in equation~\eqnCl.
\smallskip
\goodbreak
\centerline{\largebf Appendix B. Single Fluid and Other Useful Relations}
\smallskip
Above the horizon the entropy perturbation $S$ is constant, and
all perturbation quantities can be obtained from
the solution for the total density perturbation $\Delta_T$.  Combining
the total continuity and Euler equations in \eqnTotal\ yields the second
order evolution equation
$$
\left\{ {d^2 \over da^2} - {f \over a}{d \over da}
+ {1 \over a^2}\left[\left({k \over k_{eq}}\right)^2
\left( {1 - {3K \over k^2}} \right)h - g
\right] \right\} \Delta_T = \left({k \over k_{eq}}\right)^2
\left( {1 - {3K \over k^2}} \right)j S,
\Beqn\BeqnCTotal
$$
where
$$
\eqalign{
f &= {3a \over 4+3a}
        -{5 \over 2} {a \over 1+a}, \cr
g &= 2 +{9a \over 4+3a} - {a \over 2}{6+7a \over (1+a)^2}, \cr
h &= {8 \over 3} {a^2 \over (4+3a)(1+a)}, \cr
j &= {8 \over 3} {a \over (4+3a)(1+a)^2}, \cr
}
\Beqn\BeqnFunctions
$$
where recall that $a$ is normalized to unity at matter-radiation
equality.
Here we have taken the anisotropic stress $\Pi = 0$ and assumed that
the universe is in the matter or radiation dominated epoch.
The solutions to the homogeneous equation with $S=0$ are given by
$$
\eqalign{
U_A & = \left[
      a^3
     + {2 \over 9} a^2
     - {8 \over 9}a
     -{16 \over 9}
     + {16 \over 9} \sqrt{a+1}   \right]
   {1 \over a(a+ 1)}\, , \cr
U_D & = {1 \over a \sqrt{a+1}}\, ,\cr
           }
\Beqn\BeqnUAD
$$
and represent the growing
and decaying mode of adiabatic perturbations respectively.
Using Green's method,
the particular solution in the presence of a {\it constant}
entropy fluctuation $S$ becomes $\Delta_T = C_A U_A + C_D U_D + S U_I$,
where $U_I$ is given by
$$
U_I = {4 \over 15} \left( k \over k_{eq} \right)^2
	   \left( {1 - {3K \over k^2}} \right)
	   { 3a^2 + 22a + 24 + 4(4+3a)(1+a)^{1/2}
           \over (1+a)( 3a+4)[1+(1+a)^{1/2}]^4 }a^3.
\Beqn\BeqnUS
$$

After radiation becomes negligible, the both isocurvature and adiabatic
modes evolve in the same manner
$$
\ddot \Delta_T + {\dot a \over a} \dot \Delta_T = 4\pi G\rho \left({
a \over a_0 }\right)^2 \Delta_T.
\Beqn\BeqnDeltaT
$$
For pressureless perturbations, each mass shell evolves as a separate
homogeneous universe.  Since a density perturbation can be viewed
as merely a different choice of the initial time surface, the evolution of the
fractional shift in the scale factor, \ie\ the Hubble parameter $H$,
must coincide with $\Delta_T$.
It is simple to check that the
Friedman equations do indeed imply
$$
\ddot H + {\dot a \over a} \dot H = 4\pi G\rho \left({
a \over a_0 }\right)^2 H,
\Beqn\BeqnH
$$
so that one solution, the decaying mode, of equation \Bdis\BeqnDeltaT\ is
$\Delta_T \propto
H$ [\PeeblesLSS].
The growing mode $\Delta_T \propto D$
can easily be determined by writing its form as
$D \propto H G$ yielding
$$
\ddot G + \left( {\dot a \over a} + 2 {\dot H \over H} \right) \dot G = 0
\Beqn\BeqnG
$$
which can be immediately solved as [\PeeblesLSS]
$$
D(a) \propto H \int {d a \over (a H)^3}.
\Beqn\BeqnD
$$
Note that we ignore pressure contributions in $H$ [{\it c.f.}
equation~\eqnReducedH].
If the cosmological constant $\Lambda=0$, this integral can be performed
analytically
$$
D(a) \propto 1 + {3 \over x} + {3 (1+x)^{1/2} \over x^{3/2} }
	\ln [(1+x)^{1/2}-x^{1/2}]
\Beqn\BeqnGrowth
$$
where $x=(\Omega_0^{-1}-1)(a/a_0)$.
In the more general case, a numerical solution to this integral
must be employed.  Since before curvature or $\Lambda$ domination
$D \propto a$,
the full solution for $\Delta_T$,
where the universe is allowed to pass through radiation,
matter and curvature or $\Lambda$ domination,
can be simply obtained from equation \Bdis\BeqnUAD\ and \Bdis\BeqnUS,
by replacing $a$ with $D$ normalized so
that $D=a$ early on.

With the solution for $\Delta_T$ and
the definition of $S$ [equation~\eqnEntropy], all component
perturbations can be written in terms of
$\Delta_T$.
For example, in the baryonic isocurvature scenario,
$$
\Delta_b  = {1 \over 4+3a }[4S + 3(1+a)\Delta_T],
\Beqn\BeqnDeltab
$$
and
$$
\eqalign{
\Delta_\nu & = {4 \over 3} (\Delta_b - S_{b\nu}), \cr
\Delta_\gamma & = {4 \over 3} (\Delta_b - S_{b\gamma}). \cr }
\Beqn\BeqnDeltar
$$
The fact that in this model
the curvature perturbation vanishes initially when the universe is radiation
dominated allows us to
set $S_{b\nu}=S_{b\gamma}$.
The velocity and potentials can be written as
$$
\eqalign{
V_T &= -{3 \over k}{\dot a \over a}\left( {1-{3K \over k^2}} \right)^{-1}
	{{1+a} \over {4+3a}}\left[ a {d\Delta_T \over da} -
	{1 \over {1+a}} \Delta_T \right], \cr
\Psi & = - {3 \over 4} \left( {k_{eq} \over k} \right)^2
	   \left( {1 - {3K \over k^2}} \right)^{-1}
	   {1 + a \over a^2}
\Delta_T,
\cr}
\Beqn\BeqnContPois
$$
where note that constant entropy assumption
requires that all the velocities $V_i = V_T$.
The relation for the velocity may be simplified by noting that
$$
\eqalign{
\eta(a) &\approx {2\sqrt{2} \over k_{eq}} \left[ \sqrt{1+a} -1  \right] \qquad
{\rm RD/MD} \cr
&\approx {1 \over \sqrt{-K}} \cosh^{-1}\left[ 1 + {2 (1-\Omega_0) \over
\Omega_0} {a \over a_0} \right], \qquad {\rm MD/CD} \cr}
\Beqn\BeqnEta
$$
where CD denotes curvature domination with $\Lambda=0$.  For $\Lambda \ne
0$, it must be evaluated by numerical integration.   Before curvature or
$\Lambda$ domination
$$
{\dot a \over a} = {(1+a)^{1/2} \over \sqrt{2} a} k_{eq},
\Beqn\BeqnDota
$$
which can be used to explicitly evaluate \Bdis\BeqnContPois.
Finally, in Tab.~1 we list some commonly used symbols in the paper and
the equation in which they first appeared.
\bigskip
\vfill
\baselineskip =12truept \rightskip=3truepc \leftskip 3truepc
\noindent {\bf Table 1.} Commonly used symbols.  Time variables $a$,
$z$, $\eta$, and $\chi$ are often used interchangably with special
epochs listed here under scale factor $a$ entries. Component
density $\Delta_i$ and velocity $V_i$ are defined in \S IIC and D,
with $i$ as $b$ for baryons, $\gamma$ for photons, $\nu$ for
neutrinos, and $c$ for collisionless cold dark matter.
Note that $V_\gamma=\Theta_1$ ({\it see following page}).
\eject
 \vbox{ \vskip 20pt \centerline{
 \vbox{ \offinterlineskip
 \halign { \vrule#
 & \quad # \hfil& \vrule#
 & \quad # \hfil& \vrule#
 & \quad # \hfil& \vrule# \cr
 \noalign{\hrule} height2pt
 &\omit& &\omit& &\omit&
 \cr
 & Symbol~& & Definition \quad&& Equation~& \cr
 height2pt
 &\omit& &\omit& &\omit&
 \cr \noalign{\hrule}
 height2pt
 &\omit& &\omit& &\omit&
 \cr
&$\Delta_T$     &&Total~density~fluctuation
&&~\eqnDeltadef                  & \cr
 height2pt
 &\omit& &\omit& &\omit&
 \cr \noalign{\hrule}
 height2pt
 &\omit& &\omit& &\omit&
 \cr
&$\Theta$       &&CMB~temperature~fluctuation
&&~\eqnGen                       & \cr
 height2pt
 &\omit& &\omit& &\omit&
 \cr \noalign{\hrule}
 height2pt
 &\omit& &\omit& &\omit&
 \cr
&$\Theta_0$     &&CMB~monopole~fluctuation
&&~\eqnHierarchy                 & \cr
 height2pt
 &\omit& &\omit& &\omit&
 \cr \noalign{\hrule}
 height2pt
 &\omit& &\omit& &\omit&
 \cr
&$\Theta_\ell$  &&CMB~$\ell$th~multipole~fluctuation
&&~\eqnHierarchy                 & \cr
 height2pt
 &\omit& &\omit& &\omit&
 \cr \noalign{\hrule}
 height2pt
 &\omit& &\omit& &\omit&
 \cr
&$\Pi$          &&Anisotropic~stress~perturbation
&&~\eqnAniso                     & \cr
 height2pt
 &\omit& &\omit& &\omit&
 \cr \noalign{\hrule}
 height2pt
 &\omit& &\omit& &\omit&
 \cr
&$\Psi$         &&Gravitational~(Newtonian)~potential
&&~\eqnPotential                 & \cr
 height2pt
 &\omit& &\omit& &\omit&
 \cr \noalign{\hrule}
 height2pt
 &\omit& &\omit& &\omit&
 \cr
&$\Phi$         &&Gravitational~(curvature)~potential
&&~\eqnPotential                 & \cr
 height2pt
 &\omit& &\omit& &\omit&
 \cr \noalign{\hrule}
 height2pt
 &\omit& &\omit& &\omit&
 \cr
&$\Psi$         &&Gravitational~(curvature)~potential
&&~\eqnPotential                 & \cr
 height2pt
 &\omit& &\omit& &\omit&
 \cr \noalign{\hrule}
 height2pt
 &\omit& &\omit& &\omit&
 \cr
&$\eta$         &&Conformal~time
&&~\eqnFRW                       & \cr
 height2pt
 &\omit& &\omit& &\omit&
 \cr \noalign{\hrule}
 height2pt
 &\omit& &\omit& &\omit&
 \cr
&$\nu$          &&Curvature~normalized~wavenumber
&&~\eqnSachsWolfe                & \cr
 height2pt
 &\omit& &\omit& &\omit&
 \cr \noalign{\hrule}
 height2pt
 &\omit& &\omit& &\omit&
 \cr
&$\sigma_T$     &&Thomson~cross~section
&&~\eqnGen                       & \cr
 height2pt
 &\omit& &\omit& &\omit&
 \cr \noalign{\hrule}
 height2pt
 &\omit& &\omit& &\omit&
 \cr
&$\tau$         &&Thomson~optical~depth
&&~\eqnGen                       & \cr
 height2pt
 &\omit& &\omit& &\omit&
 \cr \noalign{\hrule}
 height2pt
 &\omit& &\omit& &\omit&
 \cr
&$\chi$         &&Curvature~normalized~distance
&&~\eqnSachsWolfe                & \cr
 height2pt
 &\omit& &\omit& &\omit&
 \cr \noalign{\hrule}
 height2pt
 &\omit& &\omit& &\omit&
 \cr
&${\cal~D}$     &&Diffusion~damping~factor
&&~\eqnDamp                      & \cr
 height2pt
 &\omit& &\omit& &\omit&
 \cr \noalign{\hrule}
 height2pt
 &\omit& &\omit& &\omit&
 \cr
&${\cal~G}$     &&Drag~growth~factor
&&~\eqnDragGrow                  & \cr
 height2pt
 &\omit& &\omit& &\omit&
 \cr \noalign{\hrule}
 height2pt
 &\omit& &\omit& &\omit&
 \cr
&$C_A$          &&Initial~adiabatic~spectrum
&&~\eqnGenSolution               & \cr
 height2pt
 &\omit& &\omit& &\omit&
 \cr \noalign{\hrule}
 height2pt
 &\omit& &\omit& &\omit&
 \cr
&$C_I$          &&Initial~isocurvature~spectrum
&&~\eqnGenSolution               & \cr
 height2pt
 &\omit& &\omit& &\omit&
 \cr \noalign{\hrule}
 height2pt
 &\omit& &\omit& &\omit&
 \cr
&$C_\ell$       &&Anisotropy~power~spectrum
&&~\eqnCl                        & \cr
 height2pt
 &\omit& &\omit& &\omit&
 \cr \noalign{\hrule}
 height2pt
 &\omit& &\omit& &\omit&
 \cr
&$D$            &&Pressureless~growth~factor
&&~\eqnDgrowth                   & \cr
 height2pt
 &\omit& &\omit& &\omit&
 \cr \noalign{\hrule}
 height2pt
 &\omit& &\omit& &\omit&
 \cr
&$F$            &&Gravitational~driving~force
&&~\eqnForce                     & \cr
 height2pt
 &\omit& &\omit& &\omit&
 \cr \noalign{\hrule}
 height2pt
 &\omit& &\omit& &\omit&
 \cr
&$H$            &&Hubble~parameter
&&~\eqnHubble                    & \cr
 height2pt
 &\omit& &\omit& &\omit&
 \cr \noalign{\hrule}
 height2pt
 &\omit& &\omit& &\omit&
 \cr
&$N_\ell$       &&Neutrino~$\ell$th~multipole
&&~\eqnHierarchy                 & \cr
 height2pt
 &\omit& &\omit& &\omit&
 \cr \noalign{\hrule}
 height2pt
 &\omit& &\omit& &\omit&
 \cr
&$K$            &&Curvature
&&~\eqnFRW                       & \cr
 height2pt
 &\omit& &\omit& &\omit&
 \cr \noalign{\hrule}
 height2pt
 &\omit& &\omit& &\omit&
 \cr
&$R$            &&Normalized~scale~factor~$3\rho_b/4\rho_\gamma$
&&~\eqnBaryon                    & \cr
 height2pt
 &\omit& &\omit& &\omit&
 \cr \noalign{\hrule}
 height2pt
 &\omit& &\omit& &\omit&
 \cr
&$S$            &&Entropy~fluctuation
&&~\eqnTotal                     & \cr
 height2pt
 &\omit& &\omit& &\omit&
 \cr \noalign{\hrule}
 height2pt
 &\omit& &\omit& &\omit&
 \cr
&$T$            &&Matter~transfer~function
&&~\eqnTransfer                  & \cr
 height2pt
 &\omit& &\omit& &\omit&
 \cr \noalign{\hrule}
 height2pt
 &\omit& &\omit& &\omit&
 \cr
&$V_T$          &&Total~velocity~amplitude
&&~\eqnDeltadef                  & \cr
 height2pt
 &\omit& &\omit& &\omit&
 \cr \noalign{\hrule}
 height2pt
 &\omit& &\omit& &\omit&
 \cr
&$X_\nu^\ell$   &&Radial~eigenfunction
&&~\Adis\AeqnXIso                & \cr
 height2pt
 &\omit& &\omit& &\omit&
 \cr \noalign{\hrule}
 height2pt
 &\omit& &\omit& &\omit&
 \cr
&$a$            &&Scale~factor
&&~\eqnFRW                       & \cr
 height2pt
 &\omit& &\omit& &\omit&
 \cr \noalign{\hrule}
 height2pt
 &\omit& &\omit& &\omit&
 \cr
&$a_0$          &&Present~scale~factor
&&~\eqnFRW                       & \cr
 height2pt
 &\omit& &\omit& &\omit&
 \cr \noalign{\hrule}
 height2pt
 &\omit& &\omit& &\omit&
 \cr
&$a_d$          &&Compton~drag~epoch
&&~\eqnZdrag                     & \cr
 height2pt
 &\omit& &\omit& &\omit&
 \cr \noalign{\hrule}
 height2pt
 &\omit& &\omit& &\omit&
 \cr
&$a_{eq}$       &&Equality~scale~factor
&&~\eqnFRW                       & \cr
 height2pt
 &\omit& &\omit& &\omit&
 \cr \noalign{\hrule}
 height2pt
 &\omit& &\omit& &\omit&
 \cr
&$a_i$          &&Ionization~epoch
&&~\eqnGrowLate                  & \cr
 height2pt
 &\omit& &\omit& &\omit&
 \cr \noalign{\hrule}
 height2pt
 &\omit& &\omit& &\omit&
 \cr
&$a_*$          &&Recombination~conformal~time
&&~\eqnZLS                       & \cr
 height2pt
 &\omit& &\omit& &\omit&
 \cr \noalign{\hrule}
 height2pt
 &\omit& &\omit& &\omit&
 \cr
&$c_s$          &&Photon-baryon~sound~speed
&&~\eqnSoundSpeed                & \cr
 height2pt
 &\omit& &\omit& &\omit&
 \cr \noalign{\hrule}
 height2pt
 &\omit& &\omit& &\omit&
 \cr
&$k$            &&Laplacian~wavenumber
&&~\eqnEigen                     & \cr
 height2pt
 &\omit& &\omit& &\omit&
 \cr \noalign{\hrule}
 height2pt
 &\omit& &\omit& &\omit&
 \cr
&$\tilde k$     &&Renormalized~wavenumber
&&~\eqnEigen                     & \cr
 height2pt
 &\omit& &\omit& &\omit&
 \cr \noalign{\hrule}
 height2pt
 &\omit& &\omit& &\omit&
 \cr
&$k_D$          &&Diffusion~damping~wavenumber
&&~\eqnDampLength                & \cr
 height2pt
 &\omit& &\omit& &\omit&
 \cr \noalign{\hrule}
 height2pt
 &\omit& &\omit& &\omit&
 \cr
&$k_{eq}$       &&Equality~horizon~wavenumber
&&~\eqnDeltaLS                   & \cr
 height2pt
 &\omit& &\omit& &\omit&
 \cr \noalign{\hrule}
 height2pt
 &\omit& &\omit& &\omit&
 \cr
&$\ell$         &&Multipole~number
&&~\eqnLDecomposition            & \cr
 height2pt
 &\omit& &\omit& &\omit&
 \cr \noalign{\hrule}
 height2pt
 &\omit& &\omit& &\omit&
 \cr
&$r_\theta$     &&Projection~factor
&&~\eqnRtheta                    & \cr
 height2pt
 &\omit& &\omit& &\omit&
 \cr \noalign{\hrule}
 height2pt
 &\omit& &\omit& &\omit&
 \cr
&$r_s$          &&Sound~horizon
&&~\eqnSoundHorizon              & \cr
 height2pt
 &\omit& &\omit& &\omit&
 \cr \noalign{\hrule}
 height2pt
 &\omit& &\omit& &\omit&
 \cr
&$x_e$          &&Electron~ionization~fraction
&&~\eqnGen                       & \cr
 height2pt &\omit& &\omit& &\omit&
 \cr \noalign{\hrule} }}}
 \vskip 10pt}
\eject
\centerline{\largebf Figure Captions:}
\bigskip
\noindent {\bf Figure 1.}
Large scale open isocurvature evolution ($\Omega_0=0.2,h=0.5$,
no recombination).
Perturbations, which originate
in the baryons, are transferred to the radiation as the universe
becomes more matter dominated to avoid a significant curvature
perturbation.  Nonetheless, radiation fluctuations create
total density fluctuations from feedback.  These adiabatic
fluctuations in $\Delta_T$ dominate over the original entropy
perturbation near horizon crossing $a_H$
in the matter dominated
epoch.  The single fluid approximation cannot extend after last
scattering for the photons
$a_*$, since free streaming will damp $\Delta_\gamma$
away.  After curvature domination the total density is prevented from
growing and thus leads to decay in the gravitational potential
$\Psi$.
\smallskip

\noindent {\bf Figure 2.} The total Sachs-Wolfe
effect ($\Omega_0=0.1,h=0.5$, standard recombination).
In the adiabatic case, temperature fluctuations are
enhanced in gravitational wells such that $\Theta$ and $\Psi$
cancel, yielding $\Theta_0+\Psi={\frac 1/3} \Psi$ in the
matter dominated epoch.
For the isocurvature case,
the ISW effect creates a net total of
$\Theta_0+\Psi = 2\Psi$ reflecting the anticorrelated nature
of radiation and total density fluctuations.
After last scattering at $a_*$, this SW contribution
(analytic only) collisionlessly damps
from the monopole. The rms
temperature fluctuations (numerical only) acquires
contributions
after $a_*$ from the ISW effect due to the
radiation (early) {\it and}
curvature or $\Lambda$ (late)
contributions.
These contributions are relatively more important for adiabatic
models due to the partial cancellation of $\Theta_0$ and $\Psi$
at last scattering.
Since $\Lambda$ domination can only have occurred comparatively
recently, the late ISW effect is also less important in a $\Lambda$
compared
to an open universe.
\smallskip
\noindent {\bf Figure 3.}
$\Omega_0=1$ adiabatic full photon spectrum (standard recombination).
Shown here and
in Figs.~4,6,7 is
the contribution to the anisotropy
per logarithmic $\tilde k$ and $\ell$ interval $(\Delta T/T)^2_{\ell k}$
 [equation \eqnWeight]
with equally spaced
contours up to a cut off set to best display the features in
question.  The strong correlation
between $\ell$ and $k$ merely reflects the projection of a scale
on
the last scattering surface to an angle on the sky.
At $\log\ell \simgt 2$, SW contributions fall off and
are replaced by the acoustic peaks
(saturated here).
The detailed structure can
be traced to the radial eigenfunction $X_\nu^\ell(\chi)=j_\ell(x)$
which governs
the projection and free streaming oscillations.
\smallskip

\noindent {\bf Figure 4.}
$\Lambda$ adiabatic photon spectrum ($\Omega_0=0.1,h=0.5$,
standard recombination).
Unlike the $\Omega_0=1$
case, this scenario has significant contributions from after
last scattering through the early and late
ISW effect.  (a) The early ISW effect arises if horizon
crossing is near radiation domination, and projects
onto a second ridge
which is more prominent than the SW
ridge at intermediate but not large angles.
(b) After $\Lambda$ domination, the late ISW contributions
come free streaming in from the monopole yielding a boost in
the low order multipoles for a small range in $k$, due to cancellation
with SW contributions at the largest scales and crest-trough cancellation
at smaller scales.  Scales depicted in Fig.~5a,b are marked here by
dashed lines.
\smallskip

\noindent {\bf Figure 5.}
Analytic separation of adiabatic large angle anisotropies
($\Omega_0=0.1$ $h=0.5$, standard recombination, arbitrary
normalization). Scales are chosen to match the features in
Fig.~4 and 6.
$\Lambda$ models:  (a) At the largest scales, \eg\ here
$k=10^{-4}$Mpc$^{-1}$, the
SW effect dominates over the late ISW effect due to projection.
However since the potential decays, the late ISW effect partially cancels
the SW effect
if the mode is superhorizon sized at
$\Lambda$ domination. (b) Intermediate scale peaks in Fig.~4
are due to the late ISW boost of the higher SW free streaming ridges.
Open models: (c)
The maximum scale
corresponds to the curvature radius $k = \sqrt{-K}$.
For this scale, the SW effect projects broadly in $\ell$
peaking near
$\ell \sim 10$.  For the late ISW effect,
this scale projects onto the monopole and
dipole near curvature domination
thus leaving the ISW contributions to decrease smoothly with $\ell$.
(d) At smaller scales,  corresponding to the large ridge in Fig.~6,
the late ISW effect projects onto $\ell \approx
2-10$ and completely dominates leading to a rising spectrum of
anisotropies.

\smallskip
\noindent {\bf Figure 6.} Open adiabatic photon spectrum ($\Omega_0=0.1,
h=0.5$, standard recombination).
(a) Like the $\Lambda$ case, the radiation ISW effect
contributes significantly to intermediate angle anisotropies.
(b) However, as already noted in Fig.~2, the late ISW effect
appearing at the left
is much more significant than the corresponding $\Lambda$ effect.
Thus on all angular scales,
the total ISW contribution
dominates the SW effect.  Contours curve away from the
curvature scale $\log(k*$Mpc$)=-3.8$ due to suppression of the potentials
from the Poisson equation.  Scales depicted in Fig. 5c,d are marked
here with dashed lines.

\smallskip
\noindent {\bf Figure 7.}
Open and $\Lambda$ isocurvature photon spectrum
($\Omega_0=0.1,h=0.5$, standard recombination). Unlike their
adiabatic counterparts, the potential {\it grows} in the
radiation domination era only to turn over and decay in the
curvature and $\Lambda$ dominated era.
The ISW contribution will thus smoothly match onto the SW contribution.
This has the effect of merging the SW and ISW ridges to make a
wide feature that contributes broadly in $\ell$.
For $\Lambda$ models, the radiation ISW effect completely dominates
over the $\Lambda$ ISW effect. Scales depicted in Fig.~8 are
marked here in dahsed lines.

\smallskip
\noindent {\bf Figure 8.}
Analytic separation of isocurvature large angle anisotropies ($\Omega_0=0.1$,
$h=0.5$, standard recombination, arbitrary normalization).
Scales are chosen to match
the features in Fig.~7.
In general, isocurvature models have strong early ISW contributions
which mimic and coherently boost the SW effect.
$\Lambda$
models:
(a) Notice that the shape of the SW and ISW effects are identical at
large scales.
(b) Even at the late ISW peak, the early ISW contributions are so strong
that the late contributions are never apparent unlike the
adiabatic model.
Open models:  (c) As with
$\Lambda$ models, radiation epoch contributions are
significant making the SW and ISW contributions similar for
large scales.  (d) Near the peak of the curvature ISW contribution however,
the relative contributions are similar to the adiabatic case.

\smallskip
\noindent {\bf Figure 9.} The acoustic oscillations
($\Omega_0=0.2,h=0.5$, no recombination).
The photon-baryon fluid acts like an oscillator in
a potential well.
The dipole, \ie\ the photon velocity $V_\gamma$, is
increasingly suppressed with respect to the monopole as $(1+R)^{-1/2}$,
where the $\sqrt{3}$ accounts for
the three degrees of freedom in the dipole.
Scales which reach an extrema in the monopole at
last scattering will correspond to the so-called ``Doppler peaks''
in the anisotropy spectrum.
Also displayed here is the semianalytic
approximation described in the text, which is essentially
exact.
The small difference in the numerical amplitudes of $\Phi$ and $\Psi$
is due to the anisotropic stress of the neutrinos. Whereas
the isocurvature case has $\Omega_0 = \Omega_b$, the
adiabatic model has $\Omega_b=0.06$ and a consequently smaller
$R$.

\smallskip
\noindent {\bf Figure 10.} Small scale isocurvature evolution and
photon diffusion ($\Omega_0=0.2,h=0.5$, no recombination).
At small scales gravity may be ignored, yielding pure adiabatic
oscillations.
Perturbations in the photons damp once the diffusion length grows
larger than the wavelength $k_D < k$.  Likewise the
adiabatic component of the baryon fluctuations also
damps leaving
them with the original entropy perturbation.  After diffusion,
the photons and baryons behave as separate fluids, allowing
the baryons to grow once Compton drag becomes negligible $a > a_d$.
Photon fluctuations are then
regenerated by the Doppler effect as
they diffuse across infalling baryons.
The analytic approach
for the photons in this limit apply between the drag
epoch and last scattering $a_d < a < a_*$.

\smallskip
\noindent {\bf Figure 11.} Angular scale of the ``Doppler peaks''
(standard recombination).
The physical scale of the peaks is simply related
to the sound horizon
at last scattering.  Peaks in the anisotropy today will correspond to
multiples of the angle that this scale subtends on the sky $\ell_p = \pi
r_\theta
/r_s(\eta_*)$,
as discussed in the text.
Varying $\Lambda$ or $h$
increases both the sound horizon at $\eta_*$ and the present horizon
$\eta_0$ leaving
little effect. For open models,
a given scale will correspond to a smaller angle by geodesic deviation.
This simple analytic estimate for the peak location is valid
for pure acoustic contributions and underestimates the scale of
the first peak in low $\Omega_0 h^2$
models due to neglect of the early ISW effect.

\smallskip
\noindent {\bf Figure 12.} The total ISW effect ($\Omega_0=0.1,h=0.5$,
standard recombination, $k = k_3 \times 10^{-3}$Mpc$^{-1}$).
(a) Adiabatic models.  The decay of the
potential as the scale enters the horizon due to pressure growth
suppression causes the early ISW effect which boosts scales approaching
the first acoustic oscillation.  The largest scales which enter after
radiation domination are boosted by the late ISW
effects due to the rapid expansion in open and $\Lambda$ models,
leaving a deficit at intermediate scales.
(b) Isocurvature models. Scales which enter early during
radiation domination do not grow as much due to the suppression in
the potential.  This enhances the large scale with respect to the
small.  Notice that the second adiabatic oscillation
($k_3=20$)
can be comparable to the first since the turnover in $\Phi$
occurs later.
Only at the largest scales is the distinction between open and
$\Lambda$ models manifest in the rms temperature fluctuations.

\smallskip
\noindent {\bf Figure 13.} Open isocurvature baryon evolution under
partial ionization ($\Omega_0=0.2,h=0.5$).
The baryons are released to grow in pressureless linear theory
after Compton drag becomes negligible.
Since this epoch becomes
earlier as the ionization fraction is decreased, present day fluctuations
will be larger for low $x_e$, if normalized to the
ionization independent fluctuations at large scales.
Unlike the CDM case, baryons have no
potential wells into which they might fall after the drag epoch and
the transfer function is extremely sensitive to the ionization history.

\smallskip
\noindent {\bf Figure 14.} Open isocurvature
baryon evolution in late
reionization scenario ($\Omega_0=0.2,h=0.5$).
Here the universe is suddenly reionized to $x_e=1$
at redshift $z_i$ after a transparent period $ 1000 > z > z_i$. Perturbations
are released from drag following recombination only to suffer
its effects once again between the ionization and drag
epochs  $z_i > z > z_d$.  Thus the final fluctuations will be larger
for later reionization.

\smallskip
\noindent {\bf Figure 15.} Small scale isocurvature temperature evolution
under partial reionization ($\Omega_0=0.2,h=0.5$, numerical).
If the universe stays transparent after standard recombination at
$z_* \approx 1000$, the acoustic oscillations in the photon fluid
will be frozen.
However these large fluctuations are suppressed by diffusion
damping in partially reionized models.
Although the Doppler and other diffusive effects regenerate
fluctuations
at small scales, these effects are also suppressed under
the diffusion length
(\ie\ the thickness of the last scattering surface).

\smallskip
\noindent {\bf Figure 16.} The adiabatic matter transfer function
($h=0.5$,$\Omega_b=0.01$).
The transfer function has been scaled by $(\Omega_0 h^2)^{-2}$
to compare different
$\Omega_0$ values by requiring the same initial
gravitational potential $\Phi$ (below the curvature scale).
For scale invariant $n=1$ $\Lambda$ models,
this normalization
is equivalent to that determined by large scale
anisotropies,
since the SW effect dominates
all but the lowest multipoles.
Therefore the
approximate relative
amplitude
of matter fluctuations can be directly read off from this
plot.  For open models, this is not true due to a more
significant ISW effect and curvature effects at large
scales which relate the potential to the initial power
spectrum.

\smallskip
\noindent {\bf Figure 17.} The isocurvature matter transfer function
($\Omega_0=0.2,h=0.5$, numerical).
The baryon
perturbations will have a prominent peak at the maximal Jeans
scale since perturbations grow as $D(a)$ outside this scale,
with a $k^2- 3K$ tail, and are suppressed inside it.  The
acoustic oscillations damp away in the highly ionized case
since last scattering is delayed and the diffusion length grows.
This leaves a flat small scale tail in the transfer function.
Note also that in the low ionization scenarios, the Jeans length
may not grown to its maximum matter dominated value by last scattering
leading to a smaller scale for the peak. The growth suppression
due to $\Lambda$ is less significant than that from curvature
domination.

\smallskip
\noindent {\bf Figure 18.}
The full adiabatic photon power spectrum.  The logarithmic
contributions to the anisotropy in $\ell$ and $k$
[see equation~\eqnWeight]
are plotted here.  Whereas in the $\Omega_0=1$ case
only the SW ridge and acoustic peaks are prominent ({\it top left}
and close up, {\it top right}), the $\Lambda$ and open
cases show more complicated  structure due to the ISW
effect.  Depending on the initial weightings, represented here
as $|C_A|^2 \propto \tilde k$,
certain features
may be emphasized over others.
Notice the $\Lambda$ ISW effect at low $\ell$ and intermediate
$k$ and the comparatively small open SW contribution at the foot
of the ISW ridge.

\smallskip
\noindent {\bf Figure 19.} Large angle
anisotropy dependence on the initial
power spectrum $|C_I|^2 \propto \tilde k^n$ or $|C_A|^2 \propto \tilde k^n$
for isocurvature and adiabatic scenarios respectively.
($\Omega_0=0.1$, $h=0.5$, standard recombination).
Notice that
for red spectra, geometric and/or cosmological constant effects
play a role in determining the anisotropy whereas
for very blue spectra, $\ell^2 C_\ell \propto  \ell^2$ for
all models.   Isocurvature models with $n \approx -1$ to
fit large scale structure will thus not be extremely sensitive
to open or $\Lambda$ dominated universe effects.
The normalization here is arbitrarily set at the quadrupole.

\smallskip
\noindent {\bf Figure 20.}
The full open isocurvature photon power spectrum for
$|C_I|^2 \propto \tilde k^n$.
(a) Curvature scale weighted $n = -3$.  The lack of a Poisson
cut off in the isocurvature potential makes the nature of
the open universe eigenfunctions apparent.  The projection ridge
crosses minimum eigenvalue $k = \sqrt{-K}$ (front edge)
at roughly $\ell \sim 10$ corresponding to the fact that the lowest
eigenfunction still has structure only around the curvature scale.
This leads to the cutoff to low multipoles depicted in Fig.~19.
(b) Small scale weighted $n = -1$.  The main projection ridge does
not dominate the anisotropy at the low order multipoles.  Power
from smaller physical scales (high $k$) bleeds in to boost the
anisotropy.  Thus anisotropies do not
rise as rapidly with $\ell$ as predicted from the one to one conversion
of $k$ onto $\ell$.  For this model $n_{\rm eff} \approx 2$ at large angles.

\smallskip
\noindent {\bf Figure 21.} Intermediate to small scale anisotropies.
(a) Adiabatic models.
Projection of the sound horizon at last scattering onto sky today
determines the angular scale of the ``Doppler peaks'' ({\it c.f.}
Fig.~11).  The sound horizon is the same physical scale for open
and $\Lambda$ models with fixed $\Omega_0$ but geodesic deviation makes it
correspond to a smaller angle in the open case.  Compared with the flat
case, the $\Lambda$ model has a somewhat smaller angular scale due to the
imperfect cancellation between the increase in the age of the universe
today and at last scattering. (b) Isocurvature models.
Anisotropies in the standard recombination scenario ($x_e\approx 0$
if $z < 1000$) produce far
to large fluctuations on the arcminute scale due to the steeply
rising spectrum.
Reionized models have their adiabatic fluctuations
damped out by photon diffusion and a cancellation suppressed Doppler effect.
Notice that large angle anisotropies
are immune to ionization history effects for the open case but not
for the $\Lambda$ case.  This and the difference in the damping scale
is mostly due to the projection effect.

\smallskip
\noindent {\bf Figure 22.} Isocurvature temperature power spectrum.
In this fully ionized $x_e=1$, low $\Omega_0 = 0.1$ $h=0.5$ model,
the ISW effect makes a contribution equal to and with the same
scale dependence as the cancelled Doppler (plus SW)
term (DSW).  The second order Vishniac term (V) dominates at small scales.  The
analytic approximation (solid) fails at large scales where cancellation
arguments are not applicable.

\smallskip
\noindent {\bf Figure 23.} Radial eigenfunctions of an open universe
$X_\nu^\ell(\chi)$.  (a) The isotropic $\ell = 0$ function for several
values of the wavenumber $\nu = \tilde k/\sqrt{-K}$.
The zero crossing moves out to arbitrarily large scales as
$\nu \rightarrow 0$, reflecting completeness.
However, even as this ``effective wavelength''
becomes infinite, the function retains prominent structure only
near the curvature scale $\chi$.  A random superposition of
these low $\nu$ modes cannot produce more than exponentially
decaying structure larger than the curvature scale.
(b) Low order multipoles in the asymptotic limit $\nu \rightarrow 0$.
If most power lies on the curvature scale, the $\ell$-mode
corresponding to the angle that the curvature radius subtends will
dominate the anisotropy.  The normalization is appropriate for comparing
contributions to the anisotropy $\ell(2\ell+1)C_\ell/4\pi$.
Also shown is the location of the horizon $\chi = \eta_0 \sqrt{-K}$
for several values of $\Omega_0$.  If contributions to the anisotropy
come from a sufficiently early epoch, the dominant $\ell$-mode will peak
at this value.

\bye